\documentclass[a4paper,12pt]{article}
\usepackage{soul}
\usepackage[usenames,dvipsnames]{color}
\usepackage{jheppub}
\usepackage{esvect}
\usepackage{amsmath, amssymb, slashed, epsf, color, graphicx, latexsym}
\usepackage{epsfig}
\usepackage{graphics}

\newcommand{\BR}{{\bar{R}}}
\newcommand{\Bnabla}{{\bar{\nabla}}}

\begin{document}
	
	\title{The large $D$ membrane paradigm for general four-derivative theory of gravity with a cosmological constant}
	\author[a]{Aditya Kar, }
	\author[b]{Taniya Mandal, }
	\author[c] {Arunabha Saha}
	\affiliation[a]{Indian Institute of Science Education and Research Pune,
Homi Bhabha Road, Pashan, Pune 411008, India } 
	\affiliation[b]{Indian Institute of Science Education and Research Bhopal,
	Bhopal Bypass, Bhopal 462066, India } 
	\affiliation[c]{University of Geneva, 24 quai Ernest-Ansermet, 1211 Geneve 4, Switzerland}
	\emailAdd{aditya.kar@students.iiserpune.ac.in}
	\emailAdd{tmandal@iiserb.ac.in}
	\emailAdd{arunabha.saha@unige.ch}
	
	\abstract{We find the membrane equations which describe the leading order in $1/D$ dynamics of black holes in the $D\rightarrow\infty$ limit for the most general four-derivative theory of gravity in the presence of a cosmological constant. We work up to linear order in the parameter determining the strength of the four-derivative corrections to the gravity action and hence there are no ghost modes in the theory. We find that the effective membrane equations we obtain are the  covariant version of  the membrane equations in absence of the cosmological constant. We also find the world-volume stress tensor for the membrane whose conservation gives  the membrane equations. We apply the membrane equations to predict the light quasi-normal mode spectrum of black holes and black branes in the theory of gravity under consideration. }



\maketitle
\section{Introduction}
The equations of general theory of relativity governing the dynamics of spacetime are a set of coupled partial differential equations on the metric of spacetime. Since, there are no algorithmic way to solve these equations, exact analytic solutions to these equations are extremely difficult to find and there are very few such solutions known. Of them a class of interesting solutions are those of static or stationary black holes. Even though these solutions are exact, it is next to impossible to study any non-trivial dynamical processes involving these black holes analytically. For the vacuum Einstein equations there are no free parameter which can be tuned to suitable limits to get more analytic control over the problem, so that we can gain more insight into the structure of these equations and solutions.


In the last few years starting with the work of \cite{Emparan:2013moa,Emparan:2013oza,Emparan:2013xia,Emparan:2014cia,Emparan:2014jca}, the limit of taking the spacetime dimensionality to infinity has been found to be particularly useful to get more analytic control over the dynamics of black holes. In particular, it was shown in \cite{Emparan:2014aba, Emparan:2015rva} that in the $D\rightarrow\infty$ limit there is a decoupled sector of finite number of light quasi-normal modes of black holes, which completely determine the late time small amplitude dynamics.~These modes live in a region of width of  $\mathcal{O}(1/D)$ about the horizon of the black hole. In addition there also is an infinite tower of heavy quasi-normal modes which die off exponentially fast at late times. Since there is a parametric separation  between the energies of the light and heavy modes, there must be  an autonomous system of non-linear equations determining the complete late dynamics of the black holes and which in the small amplitude limit gives rise to these light quasi-normal modes. 

The effective theory of these light modes were discovered independently in two different incarnations. In one of the approaches \cite{Emparan:2015hwa, Suzuki:2015iha} the effective equations are obtained in terms of  effective mass and momentum variable associated with the black hole. In the other approach \cite{Bhattacharyya:2015dva} the theory of light modes is described by a system of equations determining the dynamics of a co-dimension one membrane moving in asymptotic spacetime of the black hole. The linearised spectrum  about static black holes obtained from both these approaches matched with the light quasi-normal modes mentioned above. 

Since then both these approaches have been extended in diverse directions.  Both these approaches have been generalised to black holes with charge and/or in presence of cosmological constant\cite{Bhattacharyya:2015fdk, Tanabe:2015isb, Bhattacharyya:2017hpj, Bhattacharyya:2018szu, Kundu:2018dvx}. In particular the effective equations obtained in the mass and momentum variable were particularly useful in understanding the  end-point of IR instability of black strings/branes\cite{Emparan:2015gva} in flat spacetime (see also\cite{Rozali:2016yhw}). The relation between the hydrodynamic limit and the large $D$ limit have also been explicated in the  mass-momentum approach in  \cite{Emparan:2016sjk} and in the membrane picture in \cite{Bhattacharyya:2018iwt, Bhattacharyya:2019mbz}. The large $D$ formalism has been applied to obtain many interesting gravity solutions. It has been applied to find new interesting stationary black-bar solutions in \cite{Andrade:2018nsz, Andrade:2018rcx} and also to find analytic bulk solutions in $AdS$ dual to Bjorken flows \cite{Casalderrey-Solana:2018uag}. 

The equivalence between the effective equations obtained in these two different approaches have not been established in complete generality but in many different situations the equivalence between the two approaches have been established: e.g the equivalence of the effective equations for stationary and static black holes have been established in \cite{Dandekar:2017aiv} and the equivalence of the effective equations about black branes have been established in \cite{Dandekar:2016jrp}. For the purposes of this paper, we will focus our attention on the membrane picture of large $D$ black hole dynamics.

A particularly interesting observation was made in \cite{Bhattacharyya:2016nhn} where it was shown using the membrane picture  that the entropy current (determining the entropy of the dual black hole) has a positive definite divergence. This information was encoded in the second order in $1/D$ membrane equations obtained in \cite{Dandekar:2016fvw}. Hence, the membrane equations somehow know about the second law of black hole thermodynamics without it being used as an input anywhere in the formalism explicitly. The immediate natural question that one can ask is if the large $D$ membrane picture can tell something about the status of the second law of black hole thermodynamics for the most general classical theory of gravity. The status of the second law of black hole thermodynamics is not yet understood in full generality. A candidate entropy for most general higher curvature theory of gravity was found in \cite{Bhattacharjee:2015yaa} which follows second law for linearised dynamics about a stationary solution. Another candidate entropy which obeys second law for higher curvature gravity in situations which preserve spherical symmetry was obtained in \cite{Bhattacharyya:2016xfs}. The membrane approach can be used to fill some of this gap by finding a candidate entropy preserving the second law for the most general theory of gravity in the large $D$ limit. 

Keeping this goal in mind, it is imperative that the effective equations for black hole dynamics in large $D$ limit are derived in many different scenarios involving higher derivative theories of gravity to gain a lot of intuition about  the structure of the membrane equations in these theories. The large $D$ formalism to study black hole dynamics for the simplest consistent higher derivative theory of gravity namely Einstein-Gauss-Bonnet (EGB) gravity was first applied to study the quasi-normal modes of static black holes from the gravity picture in \cite{Chen:2015fuf}. The authors of this paper and other collaborators worked out the effective equations for black holes and black branes dynamics in the large $D$ limit in terms of the mass and momentum variables in \cite{Chen:2017hwm, Chen:2017rxa, Chen:2018nbh}. The large $D$ effective equations in terms of the membrane variables for the EGB gravity (to linear order in the Gauss-Bonnet (GB) parameter) was worked out  in \cite{Saha:2018elg}.

In this paper we take forward the investigation in this direction to find the membrane effective equations to leading order in $1/D$  in most general four-derivative theory of gravity in the presence of a negative cosmological constant. We will work perturbatively order-by-order in the parameter determining the strength of the four-derivative part of the gravity action and we will work with only those branches of solutions which have a nice limit to solutions of Einstein-Hilbert action. This makes sure that there are no ghost modes in the theory under consideration. We will provide all our results up to linear oder in this perturbation series. 

We will try to make this paper as self-contained as possible. Since the rest of the paper will contain many technical details involving the computation, we provide a brief summary of the results obtained here.  
\subsection{Summary of results}
We will work with the most general theory of four-derivative gravity whose action is given by 
\begin{eqnarray}
	S=\int \sqrt{-g}\Bigg(&&R+(D-1)(D-2)l+ \kappa \Bigg(a_1 R^2+ a_2 R_{AB}R^{AB}+ a_3 R_{ABCD}R^{ABCD} \Bigg)\Bigg),\nonumber\\
\end{eqnarray}
where $l$ is related to the cosmological constant $\Lambda$ by $$l=\frac{-\Lambda}{(D-1)(D-2)}.$$ $\kappa$ is the parameter specifying the strength of the four-derivative term in the gravitational action. We will show that for our purposes a scaled parameter $\tilde{ \kappa}$ related to $\kappa$ by
$$\tilde{ \kappa}=\kappa(D-4)(D-3),$$
will be of particular interest.~We show that to linear order in $\tilde{\kappa}$ the membrane equation determining the leading order in large $D$ dynamics of black holes is given by
\begin{eqnarray}\label{alleq}
&&\bigg(\frac{\nabla^2 u_M}{\mathcal{K}}+u^N K_{MN}-\frac{\nabla_M  \mathcal{K}}{\mathcal{K}}\bigg(1- \frac{\tilde{\kappa}a_3\mathcal{K}^2}{(D-3)^2}\bigg)-(u.\nabla) u_M\bigg(1+\frac{\tilde{\kappa}a_3\mathcal{K}^2}{(D-3)^2}\bigg)\bigg)\mathcal{P}^M_A\nonumber\\
&&=O(1/D),\nonumber \\
&&\text{and}\quad \nabla\cdot u=\mathcal{O}(1/D),
\end{eqnarray}
where all the derivatives and dot products in the above equation are taken w.r.t the $AdS$ spacetime to which the  black hole asymptotes to. The $AdS$ radius of the asymptotic spacetime is not completely determined in terms of the  cosmological constant $`l`$, but rather is modified in presence of the higher derivative terms in the action as
$$L_{AdS}^2=\frac{1}{l(1+\tilde{ \kappa} l a_1 )}.$$
The above scaling is true to leading order in large $D$.  The membrane is defined by its shape and a unit normalised velocity vector field $u_M$ in its world-volume. $K_{MN}$ is the extrinsic curvature of the membrane and $\mathcal{K}$ is the trace of the extrinsic curvature. $\mathcal{P}^M_A$ is the projector orthogonal to the velocity field in the world-volume of the membrane. 

We also find a  stress tensor in the membrane world-volume whose conservation gives the membrane equations mentioned above. This stress tensor is given by
\begin{eqnarray}
T_{MN}&=&\frac{\mathcal{K}}{2}\left(1+\frac{\tilde{\kappa} a_3\mathcal{K}^2}{D^2}\right)u_M n_N+\left(1-\frac{\tilde{\kappa} a_3\mathcal{K}^2}{D^2}\right)\frac{K_{MN}}{2}-\frac{\nabla_M u_N+\nabla_N u_M}{2}\nonumber \\&&-\left(u_M V_N+u_N V_M\right), \nonumber\\
\text{where,}&&\nonumber\\
\quad V_M&=& -\frac{1}{2}\left(1-\frac{\tilde{\kappa} a_3 \mathcal{K}^2}{D^2}\right)\frac{\nabla_M\mathcal{K}}{\mathcal{K}}+\frac{\tilde{\kappa} a_3 \mathcal{K}^2}{D^2}(u.K)_M-\frac{\tilde{\kappa} a_3 \mathcal{K}^2}{2D^2} u\cdot \nabla u_M \nonumber \\ &&+\left(1+\frac{\tilde{\kappa} a_3 \mathcal{K}^2}{D^2}\right)\frac{\nabla^2 u_M}{\mathcal{K}}.
\end{eqnarray}
 As an application of the membrane equations, we compute the spectrum of linearised fluctuations about the static spherical membrane and the static flat membrane. These spectra give predictions for the light quasi-normal modes about static black hole and black branes in the gravity theory under consideration. The scalar and vector frequencies of the spectrum of quasi-normal mode about the static black hole are given by
\begin{eqnarray}
w^s&=&\pm \left(\frac{1}{\sqrt{j(1+l)-1}}\Big(\big(1-a_3\tilde{\kappa}(1+l)\big)(j-1)+j l \Big)+\frac{a_1\tilde{\kappa}j l^2}{2\sqrt{j(1+l)-1}}\right)\nonumber \\
&&+i (1-a_3 \tilde{\kappa}(1+l))(1-j),
\end{eqnarray}
and
\begin{equation}
w_v=i (1-a_3 \tilde{\kappa}(1+l))(1-j),
\end{equation}
where $j$ represents the angular momentum number of spherical harmonics of the background  $SO(D-2)$ isometry group.
The frequency of scalar quasi-normal mode about the black brane are obtained for two different scalings in $D$ for the momenta. The quasi-normal mode for momenta of the order of $\mathcal{O}(1)$ are given by
\begin{eqnarray}
&&w_s=\pm\left(1+\frac{\tilde{ \kappa}}{2} a_1 l\right)\frac{k}{\sqrt{D}},\nonumber\\
&&w_v=\mathcal{O}(1/D).
\end{eqnarray}
The spectra when the momentum of the modes are $\mathcal{O}(\sqrt{D})$ are given by 
\begin{eqnarray}
w_s &=& \pm\left(1+\tilde{\kappa}a_1 \frac{l}{2}\right)\frac{k}{\sqrt{D}}-i \left(1-\tilde{\kappa}a_3 l\right)\frac{k^2}{l D}\quad \text{where} \quad  k=\sqrt{k^ak_a}, \nonumber\\
w_v &=& -i\frac{k^2}{l D}\left(1-\tilde{\kappa}a_3l\right).
\end{eqnarray}
In all the results for quasi-normal mode about black branes mentioned above $k$ denotes the amplitude of the momenta along the planar directions.

\section{Most general four-derivative theory of gravity}
The most general action for a theory of  four-derivative gravity in the presence of a negative cosmological constant can be written as
\begin{eqnarray}
S=\int \sqrt{-g}\Bigg(&&R+(D-1)(D-2)l+ \kappa \Bigg(a_1 R^2+ a_2 R_{AB}R^{AB}+ a_3 R_{ABCD}R^{ABCD}\nonumber\\&&+a_4 \nabla^2 R +a_5 \nabla^A\nabla^BR_{AB}\Bigg) \Bigg).
\end{eqnarray}
Here, $\kappa$ is a constant with the dimension of the square of length and $a_1\cdots a_5$ are the arbitrary dimensionless coefficients determining the relative strength of the five possible terms with a total of four derivatives acting on the metric. We take the point of view that the four-derivative terms are corrections over the Einstein-Hilbert action and we will be only looking for solutions that can be perturbatively connected to the solutions of the Einstein-Hilbert action. In this paper, we will provide all results up to  linear order in $\kappa$ only. We also take the length scale corresponding to the cosmological constant given by $l^{-1/2}$ to be an $\mathcal{O}(1)$ number in any spacetime dimension. 

Since the terms appearing with coefficients $a_4,a_5$ can be recast as the divergence of a vector, they can be integrated by parts and converted to boundary terms. So, these terms do not contribute to the bulk gravity equation of motion and we will drop these terms from the action from now on. The action that we will work with in this paper is given by\footnote{The dual hydrodynamics of this action has been studied in \cite{Kats:2007mq, Banerjee:2009wg}.}
\begin{eqnarray}
S=\int \sqrt{-g}\Bigg(&&R+(D-1)(D-2)l+ \kappa \Bigg(a_1 R^2+ a_2 R_{AB}R^{AB}+ a_3 R_{ABCD}R^{ABCD}\Bigg) \Bigg).\nonumber\\
\end{eqnarray}
This action reduces to the Einstein-Gauss-Bonnet action for the particular values of the coefficients  $ a_1=1, a_2=-4$ and $a_3=1$. The gravity equation of motion corresponding to this action is given by

\begin{eqnarray}\label{eom}
 &&R_{AB}-\frac{1}{2}g_{AB}(R+l(D-1)(D-2) ) +\kappa\Bigg(-\frac{g_{AB}}{2}(a_1 R^2+a_2 R_{CD}R^{CD} \nonumber \\ &&+a_3 R_{CDEF}R^{CDEF}) +2 a_1 R R_{AB}-(2a_1+a_2+2a_3)\nabla_A \nabla_B R +(2a_1+\frac{1}{2}a_2)\Box R g_{AB}\nonumber\\&&
+(2 a_2+4 a_3)R_{ACBD}R^{CD}+(a_2+4a_3)\Box R_{AB}+2 a_3 R_{ACDE}{R_{B}}^{CDE}-4 a_3 R_{AC}{R^{C}}_{B}\Bigg)=0.\nonumber\\
 \end{eqnarray}
  It is easy to see from the above equation of motion that for the combination of the coefficients for which the action reduces to the 
 Einstein-Gauss-Bonnet action, there are no terms in the equation with more than two derivatives acting on the metric\footnote{Though there are terms in the EGB equation which are products of terms with two derivatives acting on the metric.}. Hence, for generic choice of coefficients there can be ghost modes (modes which are ill-behaved in the sense that they can travel faster than the speed of light) in the theory. Since, we will be working perturbatively in the parameter $\tilde{\kappa}$ about solutions of two-derivative gravity and so we will not encounter these ghost modes.


\section{The leading order ansatz from static black hole}
The aim of this work is to find dynamical solutions with horizons of the gravity equations \eqref{eom} in a series in  $1/D$. In the limit of $D\rightarrow\infty$ each new term in the series adds to the accuracy of the solution in the $D\rightarrow\infty$ limit\footnote{One should not expect though that a very large number of terms in the series will give an accurate prediction of the black hole dynamics at reasonably low values of $D$. The series is most likely asymptotic as it definitely fails to capture the dynamics of the modes that were heavy in the large $D$ limit.}. To start this procedure we need to find a solution which to leading order in large $D$ solves for the gravity equation and also has a ``horizon''. We will follow the algorithm mentioned in \cite{Bhattacharyya:2015dva, Bhattacharyya:2015fdk, Dandekar:2016fvw, Bhattacharyya:2017hpj, Bhattacharyya:2018szu, Saha:2018elg} of using the metric of static black holes in Kerr-Schild coordinates to find the starting ansatz solution.

 
The metric of the  static spherically symmetric black hole which asymptotes to $AdS$ spacetime and solves the gravity equation \eqref{eom} (to linear order in $\kappa$) is given by
 \begin{equation}
 ds^2 =-f(r)d\tilde{t}^2+\frac{dr^2}{f(r)}+r^2 d{\Omega^2}_{D-2},
 \end{equation}
 where 
 \begin{eqnarray}
 \nonumber f(r) &=& 1+l r^2+ \tilde{\kappa}\mathcal{\tilde{A}}l^2r^2-g(r),\\
 \nonumber	g(r)&=& \Big(\frac{r_h}{r}\Big)^{D-3}(1+l {r_h}^2+\tilde{ \kappa}\mathcal{\tilde{A}}l^2{r_h}^2)\\
&&+\tilde{\kappa}a_3 \frac{{r_h}^{D-5}}{r^{D-3}}\Big(1-\Big(\frac{r_h}{r}\Big)^{D-1}\Big)(1+l {r_h}^2+\tilde{\kappa}\tilde{\mathcal{A}}l^2 r_h^2)^2,\nonumber\\
\tilde{\mathcal{A}}&=&\frac{1}{(D-3)(D-2)}\Big((D-1)(D a_1+a_2)+2a_3\Big),\nonumber\\
 \tilde{\kappa}&=&(D-4)(D-3)\kappa.
 \end{eqnarray}
 We do a coordinate transformation under which we replace the length element $d\tilde{t}$ with
 \begin{equation}
 d\tilde{t}=\frac{dv}{\sqrt{1+l r^2+\tilde{\kappa}\tilde{\mathcal{A}}l^2 r^2}}-\frac{dr}{f(r)},
 \end{equation}
 and find
 \begin{equation}
 ds^2 = -\frac{f(r)}{1+l r^2+\tilde{\kappa}\tilde{\mathcal{A}} l^2 r^2} dv^2+\frac{2dv dr}{\sqrt{1+l r^2+\tilde{\kappa}\tilde{\mathcal{A}} l^2 r^2}}+r^2 d\Omega^2.
  \end{equation}
 Subsequently replacaing $dv$ with
 \begin{equation}
 dv=\sqrt{1+l r^2+\tilde{\kappa}\tilde{\mathcal{A}}l^2 r^2}dt+\frac{dr}{{\sqrt{1+l r^2+\tilde{\kappa}\tilde{\mathcal{A}}l^2 r^2}}}
 \end{equation}
we arrive at the metric
 \begin{eqnarray}\label{Kerr-Schild}
ds^2&=&ds_{AdS}^2+\frac{g(r)}{1+l r^2+\tilde{\kappa}\tilde{\mathcal{A}} l^2 r^2}\left(\sqrt{1+l r^2+\tilde{\kappa}\tilde{\mathcal{A}} l^2 r^2}dt+\frac{dr}{\sqrt{1+l r^2+\tilde{\kappa}\tilde{\mathcal{A}} l^2 r^2}}\right)^2 ,\nonumber \\
 \end{eqnarray}
 where
 \begin{eqnarray}
 \frac{g(r)}{1+l r^2+\tilde{\kappa}\tilde{\mathcal{A}} l^2 r^2} &=& \left(\frac{r_h}{r}\right)^{D-3}\left(1+\frac{\tilde{ \kappa}a_3(1+l r^2+\tilde{\kappa}\tilde{\mathcal{A}} l^2 r^2)}{r_h^2}\right)\nonumber\\&&-\frac{\tilde{ \kappa}a_3(1+l r^2+\tilde{\kappa}\tilde{\mathcal{A}} l^2 r^2)}{r_h^2}\left(\frac{r_h}{r}\right)^{2(D-2)},\nonumber\\
 \end{eqnarray}
 and 
 \begin{eqnarray}
ds_{AdS}^2 &=& -dt^2(1+l r^2+\tilde{\kappa}\tilde{\mathcal{A}} l^2 r^2)+\frac{dr^2}{1+l r^2+\tilde{\kappa}\tilde{\mathcal{A}} l^2 r^2}+r^2d\Omega_{D-2}^2 ,
 \end{eqnarray}
 is the $AdS$ spacetime to which the black hole asymptotes to. The $AdS$ radius of this spacetime is given by $\frac{1}{\sqrt{\tilde{l}}}$ where, $\tilde{l}=l(1+\tilde{\kappa}\tilde{\mathcal{A}}l)$.
 The above metric is the Kerr-Schild form of the metric of the static black hole in asymptotic $AdS$ spacetime. 
 
The trace of the extrinsic curvature of the surface $r=r_h$ embedded in the asymptotic $AdS$ spacetime defined above is given by
\begin{equation}
\mathcal{K}=\frac{D}{r}\sqrt{1+l r^2+\tilde{\kappa}\tilde{\mathcal{A}} l^2 r^2}.
\end{equation}
Using this, the Kerr-Schild metric \eqref{Kerr-Schild} can be written in a more covariant form as
\begin{equation}\label{ansatz}
ds^2=ds_{AdS}^2+\left(\left(1+\tilde{ \kappa}a_3\frac{\mathcal{K}^2}{D^2}\right)\psi^{-D}-\tilde{ \kappa}a_3\frac{\mathcal{K}^2}{D^2}\psi^{-2D}\right)\left(O_Mdx^M\right)^2,
\end{equation}
where,
\begin{eqnarray}
\psi&=&\frac{r}{r_h},\nonumber\\
O_Mdx^M&=&n_M dx^M-u_M dx^M,\nonumber\\
n_Mdx^M&=&\frac{dr}{\sqrt{1+l r^2+\tilde{\kappa}\tilde{\mathcal{A}} l^2 r^2}},\nonumber\\
u_mdx^m&=&-dt\sqrt{1+l r^2+\tilde{\kappa}\tilde{\mathcal{A}} l^2 r^2},
\end{eqnarray}
 and as already mentioned, $ds^2_{AdS}$ is the metric of the $AdS$ spacetime whose length scale is $\frac{1}{\sqrt{\tilde{l}}}$. 
 
We elevate $\psi$ and $u$ to the status of analytic functions of the spacetime coordinates with $\mathcal{O}(1)$ derivatives while keeping the following properties intact
\begin{equation}
n=\frac{d\psi}{\sqrt{d\psi.d\psi}}\quad \text{so that} \quad n\cdot n=1, \quad u\cdot u=-1, \quad u\cdot n=0.
\end{equation}
Also, $\psi$ and $u$ are chosen to be functions of spacetime coordinates such that the metric always has a large $SO(D-p-3)$ isometry in the large $D$ limit with $p$ being $\mathcal{O}(1)$. 

When $\psi\ge1$ and $\psi-1\sim\mathcal{O}(1)$, then $\psi^{-D}\sim 0$ and the metric reduces to asymptotic $AdS$ which is obviously a solution to the equation of motion. 

Of primary interest to us will be the near-horizon ``membrane region''  where, $\psi-1\sim \frac{R}{D}$. To leading order in large $D$ then $\psi\sim e^{-R}$ and following the arguments in \cite{Bhattacharyya:2017hpj,Bhattacharyya:2018szu} it can be shown that the metric in this region solves the equation of motion provided $$\nabla\cdot u=\mathcal{O}(1/D).$$
From discussions in subsequent sections it will be clear that the above constraint to leading order in large $D$ will imply that the velocity field does not  point in the direction of the radius of the isometry direction. Once this is satisfied the metric in the membrane region to leading order maps to the metric of static black hole to leading order and hence obviously solves for the gravity equations.

 The region of spacetime where $\psi\ll1$ is in the interior of the ``horizon'' and we will be able to maintain this at subsequent orders in the $1/D$ perturbation series. Since this region is causally disconnected from the perspective of the observer outside the horizon it plays no role in the physics in that region and hence we will not worry about this here. 
 
Thus the metric \eqref{ansatz} can be taken to be the leading order ansatz metric for our $1/D$ perturbation theory as it solves the gravity equations to leading order in the regions of interest. 
\section{The dual membrane, its data and choice of auxiliary conditions }
The ansatz metric  \eqref{ansatz} can be thought of as being expressed in terms of data of a membrane propagating in the asymptotic $AdS$ spacetime of the black hole. The data defining the membrane comprise of its shape defined by the surface $\psi=1$ (which specifies the horizon of the spacetime to leading order in large $D$) and a unit-normalised time-like velocity field $u^\mu$ defined in the world-volume of the membrane. The fact that the velocity data is defined only in the world-volume can be inferred from the orthonormality of the velocity and the normal vector to the surface $\psi=1$  in \eqref{ansatz}. The large $D$ limit will allow us to express further corrections to the metric in a systematic $1/D$ expansion, in terms of higher derivatives of the shape and velocity data of the membrane. This is the reason we call it the ``large $D$ membrane paradigm". 

As explained above, we will be interested in finding the metric corrections in the `membrane region' of width of the order $\mathcal{O}(1/D)$ about the surface $\psi=1$. To get local expressions of the metric in terms of the data of the membrane located at $\psi=1$ we need to define the extension of  the shape and velocity data off the membrane surface. The extension of the shape data of the surface is related to the choice of the family of surfaces of which the surface $\psi=1$ is a particular member  e.g. this unique surface can be part of two (of many) distinct family of surfaces say $\psi^2-\psi=0$ or $\psi=1+f(\psi)$, where, $f(\psi=1)=0$. The choice of the extension of the velocity vector of the membrane surface may be thought of as the world-volume velocity fields defined on each member of the family of surfaces that one chooses. There are many different choices of doing this and the expression of the metric corrections will depend on these choices, but it is clear from the procedure that the physics content of these different metric corrections in terms of the actual data on the membrane surface remains unchanged. Following earlier works, we will keep calling this choice the choice of auxiliary conditions of the membrane data and we will make the following choice used in \cite{Bhattacharyya:2015dva, Dandekar:2016fvw, Saha:2018elg}\footnote{A set of particularly nice auxiliary conditions were used in \cite{Bhattacharyya:2017hpj, Bhattacharyya:2018szu} which gave rise to no metric corrections at first subleading order. We prefer using the auxiliary conditions mentioned above as it is much simpler to pull back the data to the membrane surface using it.}, namely,   
\begin{eqnarray}
(n.\nabla)n &=&0\label{metlift},\\
(n.\nabla)u &=&0\label{ulift}.
\end{eqnarray}

The above auxiliary conditions can be physically thought of as the lifting of  the shape data by demanding that the normal vector at each point follows its geodesic and the velocity vector is parallel transported along this geodesic. 

\section{The effective metric, dilaton field and the effective gravity equations}\label{iso}

In $D$ spacetime dimension, the number of independent components of the metric and the number of gravity equations are $\frac{D(D+1)}{2}$. Clearly, in the large $D$ limit, both the number of equations and variables are infinite. We solve the problem of correcting the metric at subleading order in $1/D$  using computers and so we will use the methods used in \cite{Bhattacharyya:2015dva, Bhattacharyya:2015fdk, Dandekar:2016fvw, Saha:2018elg} to reduce the infinite number of gravity equations to a finite number effective equations using the fact that the spacetime under considerations preserves a large $SO(D-p-3)$ isometry . 
A convenient way to parametrise spacetime that preserves $SO(D-p-3)$ isometry is to write the metric as 
\begin{equation}\label{spli_metric}
ds^2 = g_{\mu\nu}(x)dx^\mu dx^\nu +e^{\phi(x)}d\Omega_d, \quad d=D-p-3, \quad \mu,\nu=0,\cdots ,p+2.
\end{equation}
Here $g_{\mu\nu}$ is an effective metric in the $p+3$ directions where, non-trivial dynamics takes place and we also have an effective dilaton field $\phi$ determining the non-trivial warping of the isometry direction of the full spacetime. Both the effective metric and the dilaton fields depend only on the $p+3$ dimensional coordinate $x^\mu$. 

Due to symmetry properties of the spacetime the gravity equations of motion split in two distinct classes: the first one having components only along the $p+3$  $x^\mu$ directions and the second class having components only along the $d$  dimensional spherical directions. Also due to symmetry considerations all the $i,j$ equations are proportional to the unit sphere metric $\Omega_{ij}$. The dimension of the isometry group $d$ appears as a parameter in the equations. 

The component of the gravity equations in the $p+3$ directions are given by
\begin{eqnarray}\label{tensorem}
 	E_{\alpha\beta}=&& R_{\alpha\beta}-\frac{ g_{\alpha\beta}}{2}\left(R\ +l(D-1)(D-2) \right) -\frac{g_{\alpha\beta} }{2}\kappa(a_1 R^2+a_2 R_{AB}R^{AB}+a_3 R_{ABCD}R^{ABCD}) \nonumber\\
 	&& \nonumber +2 \kappa a_1 R R_{\alpha\beta}-\kappa(2a_1+a_2+2a_3)\nabla_\alpha \nabla_\beta R 
 	 \nonumber +\kappa(2a_1+\frac{1}{2}a_2)\Box R \ g_{\alpha\beta}-4\kappa a_3 R_{\alpha C}{R^{C}}_{\beta}
 	\nonumber \\
 	 &&\nonumber +\kappa(2 a_2+4 a_3)R_{\alpha C\beta D}R^{CD}+\kappa(a_2+4a_3)\Box R_{\alpha\beta}
 	 +2\kappa a_3R_{\alpha CDE}{R_{\beta}}^{CDE}=0,\nonumber
 	 \end{eqnarray}
where,
\begin{equation}
\Box  =\nabla_A\nabla^A=\nabla_\alpha\nabla^\alpha+\nabla_i\nabla^i=\nabla^2, \nonumber
\end{equation}
is the Laplacian w.r.t the full metric of spacetime. The quantities in this equation evaluate in terms of the effective metric of the reduced spacetime and the dilaton field to the following expressions
\begin{eqnarray}\label{tensor_eff_eqn}
&&R_{\alpha\beta}=\BR_{\alpha\beta}-\frac{d}{2}\Bnabla_{\alpha}\Bnabla_{\beta}\phi-\frac{d}{4}\Bnabla_{\alpha}\phi\Bnabla_{\beta}\phi,\nonumber\\
&& R=\BR-d\Bnabla^2\phi-\frac{d(d+1)}{4}\Bnabla_{\alpha}\phi\Bnabla^{\alpha}\phi+d(d-1)e^{-\phi},\nonumber\\
&& R_{ABCD}R^{ABCD}=\BR_{\alpha\beta\gamma\delta}\BR^{\alpha\beta\gamma\delta}+4d\left(\frac{1}{2}\Bnabla_{\nu}\Bnabla_{\gamma}\phi+\frac{1}{4}\Bnabla_{\nu}\phi\Bnabla_{\gamma}\phi\right)\left(\frac{1}{2}\Bnabla^{\nu}\Bnabla^{\gamma}\phi+\frac{1}{4}\Bnabla^{\nu}\phi\Bnabla^{\gamma}\phi\right)\nonumber\\
&&~~~~~~~~~~~~~~~~~~~~~~+2e^{-2\phi}d(d-1)\left(\frac{1}{4}\Bnabla_{\mu}\phi\Bnabla^{\mu}\phi e^{\phi}-1\right)^2,\nonumber\\
&&R_{AB}R^{AB}=\left(\BR_{\alpha\beta}-\frac{d}{2}\Bnabla_{\alpha}\Bnabla_{\beta}\phi-\frac{d}{4}\Bnabla_{\alpha}\phi\Bnabla_{\beta}\phi\right)\left(\BR^{\alpha\beta}-\frac{d}{2}\Bnabla^{\alpha}\Bnabla^{\beta}\phi-\frac{d}{4}\Bnabla^{\alpha}\phi\Bnabla^{\beta}\phi\right)\nonumber\\
&&~~~~~~~~~~~~~~~~+d\left((d-1)e^{-\phi}-\frac{1}{2}\Bnabla^2\phi-\frac{d}{4}\Bnabla_{\mu}\phi\Bnabla^{\mu}\phi\right),\nonumber\\
&& R_{\mu}^AR_{A\nu}=g^{\gamma\delta}\left(\BR_{\mu\gamma}-\frac{d}{2}\Bnabla_{\mu}\Bnabla_{\gamma}\phi-\frac{d}{4}\Bnabla_{\mu}\phi\Bnabla_{\gamma}\phi\right)\left(\BR_{\delta\nu}-\frac{d}{2}\Bnabla_{\delta}\Bnabla_{\nu}\phi-\frac{d}{4}\Bnabla_{\delta}\phi\Bnabla_{\nu}\phi\right),\nonumber\\
&&R^{AB}R_{A\mu\nu B}=g^{\beta\delta}\left(\BR_{\beta\gamma}-\frac{d}{2}\Bnabla_{\beta}\Bnabla_{\gamma}\phi-\frac{d}{4}\Bnabla_{\beta}\phi\Bnabla_{\gamma}\phi\right)\BR^{\gamma}_{\mu\nu\delta}\nonumber\\
&&~~~~~~~~~~~~~~~~~~+d\left((d-1)e^{-\phi}-\frac{1}{2}\Bnabla^2\phi-\frac{d}{4}\Bnabla_{\mu}\phi\Bnabla^{\mu}\phi\right)\left(\frac{1}{2}\Bnabla_{\mu}\Bnabla_{\nu}\phi+\frac{1}{4}\Bnabla_{\mu}\phi\Bnabla_{\nu}\phi\right),\nonumber\\
&&R_{\mu ABC}R_{\nu}^{ABC}=\BR_{\mu\alpha\beta\gamma}\BR_{\nu}^{\alpha\beta\gamma}+2d\left(\frac{1}{2}\Bnabla_{\mu}\Bnabla_{\eta}\phi+\frac{1}{4}\Bnabla_{\mu}\phi\Bnabla_{\eta}\phi\right)\left(\frac{1}{2}\Bnabla_{\nu}\Bnabla^{\eta}\phi+\frac{1}{4}\Bnabla_{\nu}\phi\Bnabla^{\eta}\phi\right),\nonumber
\end{eqnarray}
\begin{eqnarray}
&& \nabla_\mu \nabla_\nu R  =  \bar{\nabla}_\mu \bar{\nabla}_\nu \bar{R} - d \bar{\nabla}_\mu \bar{\nabla}_\nu \bar{\nabla}^2 \phi -\frac{d(d+1)}{2}(\bar{\nabla}_\mu \bar{\nabla}^\alpha \phi \bar{\nabla}_\nu \bar{\nabla}_\alpha \phi + \bar{\nabla}^\alpha \phi \bar{\nabla}_\mu \bar{\nabla}_\nu \bar{\nabla}_\alpha \phi)\nonumber \\
 && ~~~~~~~~~~~~~+d(d-1)e^{-\phi}(\bar{\nabla}_\mu \phi \bar{\nabla}_\nu \phi -\bar{\nabla}_\mu \bar{\nabla}_\nu \phi),\nonumber \\
&& \nabla^2 R \ g_{\mu \nu} =g_{\mu \nu}\bigg(\bar{\nabla}^2 \bar{R}-d\bar{\nabla}^2(\bar{\nabla}^2\phi)-\frac{d(d+1)}{2}\bigg(\bar{\nabla}^\alpha\phi\bar{\nabla}^2\bar{\nabla}_\alpha\phi+\bar{\nabla}^\alpha\bar{\nabla}^\beta\phi\bar{\nabla}_\alpha\bar{\nabla}_\beta\phi\bigg)\nonumber\\
 &&~~~~~~~~~~~~~ +d(d-1)e^{-\phi}\bigg(\bar{\nabla}_\alpha\phi\bar{\nabla}^\alpha\phi-\bar{\nabla}^2\phi\bigg)+\frac{d}{2}\bar{\nabla}_{\lambda}\phi\bigg(\bar{\nabla}^\lambda\bar{R}-d\bar{\nabla}^{\lambda}\bar{\nabla^2}\phi \nonumber\\
 &&~~~~~~~~~~~~~ -\frac{d(d+1)}{2}\bar{\nabla}_{\alpha}\phi\bar{\nabla}^{\lambda}\bar{\nabla}^{\alpha}\phi  -d(d-1)e^{-\phi}\bar{\nabla}^\lambda\phi\bigg) \bigg),\nonumber\\
&& \nabla^2R_{\mu\nu}=\bar{\nabla}^2\bar{R}_{\mu\nu}-\frac{d}{2}\bar{\nabla}^2\bar{\nabla}_\mu\bar{\nabla}_\nu\phi-\frac{d}{2}\bar{\nabla}_\alpha\bar{\nabla}_\mu\phi\bar{\nabla}^\alpha\bar{\nabla}_\nu\phi-\frac{d}{4}\bigg(\bar{\nabla}_\mu\phi\bar{\nabla}^2\bar{\nabla}_\nu\phi  \nonumber\\
 &&~~~~~~~~~~~ +\bar{\nabla}_\nu\phi\bar{\nabla}^2\bar{\nabla}_\mu\phi\bigg)+\frac{d}{4}\bigg(2\bar{\nabla}_\lambda\phi\bar{\nabla}^\lambda\bar{R}_{\mu\nu}-d\bar{\nabla}_\lambda\phi\bar{\nabla}^\lambda\bar{\nabla}_\mu\bar{\nabla}_\nu\phi-\bar{\nabla}^\lambda\phi(\bar{\nabla}_\mu\phi\bar{R}_{\lambda\nu}\nonumber\\
 &&~~~~~~~~~~~ +\bar{\nabla}_{\nu}\phi\bar{R}_{\lambda\mu})\bigg)+\frac{d}{4}\bar{\nabla}_{\mu}\phi\bar{\nabla}_{\nu}\phi \bigg(2(d-1)e^{-\phi} -\bar{\nabla}^2\phi\bigg).
\end{eqnarray}
Whereas the equations in spherical directions are given by
	
 \begin{eqnarray}\label{scalem}
 	\nonumber R_{ij}-\frac{1}{2}R\  g_{ij}-\frac{l}{2}(D-1)(D-2) g_{ij} -\frac{1}{2}\kappa(a_1 R^2+a_2 R_{AB}R^{AB}+a_3 R_{ABCD}R^{ABCD}) g_{ij} \\ 
 	 \nonumber +2 \kappa a_1 R R_{ij}-\kappa(2a_1+a_2+2a_3)\nabla_i \nabla_j R \\
 	 \nonumber +\kappa(2a_1+\frac{1}{2}a_2)\Box R \ g_{ij}
 	 +\kappa(2 a_2+4 a_3)R_{i Cj D}R^{CD}+\kappa(a_2+4a_3)\Box R_{ij}\\
 	 \nonumber +2\kappa a_3 R_{i CDE}{R_{j}}^{CDE}-4\kappa a_3 R_{i C}{R^{C}}_{j}=0,\nonumber
 	 \end{eqnarray}
where,
\begin{eqnarray}\label{scalar_eff_eqn}
&&R_{ij}=e^{\phi}\Omega_{ij}\left((d-1)e^{-\phi}-\frac{1}{2}\Bnabla^2\phi-\frac{d}{4}\Bnabla_{\mu}\phi\Bnabla^{\mu}\phi\right),\nonumber\\
&& R=\BR-d\Bnabla^2\phi-\frac{d(d+1)}{4}\Bnabla_{\mu}\phi\Bnabla^{\mu}\phi+d(d-1)e^{-\phi},\nonumber\\
&& R_{i}^AR_{Aj}=e^{\phi}\left((d-1)e^{-\phi}-\frac{1}{2}\Bnabla^2\phi-\frac{d}{4}\Bnabla_{\mu}\phi\Bnabla^{\mu}\phi\right)^2\Omega_{ij},\nonumber\\
&& R^{AB}R_{AijB}=\Bigg(e^{\phi}\left(\BR^{\mu\nu}-\frac{d}{2}\Bnabla^{\mu}\Bnabla^{\nu}\phi-\frac{d}{4}\Bnabla^{\mu}\phi\Bnabla^{\nu}\phi\right)\left(\frac{1}{2}\Bnabla_{\mu}\Bnabla_{\nu}\phi+\frac{1}{4}\Bnabla_{\mu}\phi\Bnabla_{\nu}\phi\right)\nonumber\\
&&~~~~~~~~~~~~~~~~~~+(d-1)\left((d-1)e^{-\phi}-\frac{1}{2}\Bnabla^2\phi-\frac{d}{4}\Bnabla_{\mu}\phi\Bnabla^{\mu}\phi\right)\left(\frac{1}{4}\Bnabla^{\mu}\Bnabla_{\mu}e^{\phi}-1\right)\Bigg)\Omega_{ij},\nonumber\\
&&R_{i ABC}R_{j}^{ABC}=2e^{\phi}\Omega_{ij}\left(\frac{1}{2}\Bnabla_{\alpha}\Bnabla_{\beta}\phi+\frac{1}{4}\Bnabla_{\alpha}\phi\Bnabla_{\beta}\phi\right)\left(\frac{1}{2}\Bnabla^{\alpha}\Bnabla^{\beta}\phi+\frac{1}{4}\Bnabla^{\alpha}\phi\Bnabla^{\beta}\phi\right)\nonumber\\
&&~~~~~~~~~~~~~~~~~~~~+2(d-1) e^{-\phi}\Omega_{ij}\left(\frac{e^{\phi}}{4}\nabla_{\alpha}\phi\nabla^{\phi}-1\right)^2,\nonumber
\end{eqnarray}
\begin{eqnarray}
 &&\nabla_i \nabla_j R =  \frac{g_{ij}}{2}\bar{\nabla}_{\lambda}\phi\bigg(\bar{\nabla}^{\lambda}\bar{R}-d\bar{\nabla}^{\lambda}\bar{\nabla^2}\phi-\frac{d(d+1)}{2}\bar{\nabla}_{\alpha}\phi\bar{\nabla}^{\lambda}\bar{\nabla}^{\alpha}\phi-d(d-1)e^{-\phi}\bar{\nabla^\lambda}\phi\bigg), \nonumber\\
&&  \nabla^2 R \ g_{ij} = g_{ij}\bigg(\bar{\nabla}^2 \bar{R}-d\bar{\nabla}^2(\bar{\nabla}^2\phi)-\frac{d(d+1)}{2}\bigg(\bar{\nabla}^\alpha\phi\bar{\nabla}^2\bar{\nabla}_\alpha\phi+\bar{\nabla}^\alpha\bar{\nabla}^\beta\phi\bar{\nabla}_\alpha\bar{\nabla}_\beta\phi\bigg)\nonumber\\ 
&&~~~~~~~~~~~~~~~ +d(d-1)e^{-\phi}\bigg(\bar{\nabla}_\alpha\phi\bar{\nabla}^\alpha\phi-\bar{\nabla}^2\phi\bigg)+\frac{d}{2}\bar{\nabla}_{\lambda}\phi\bigg(\bar{\nabla}^\lambda\bar{R}-d\bar{\nabla}^{\lambda}\bar{\nabla^2}\phi-\nonumber\\&&~~~~~~~~~~~~~~~~ \frac{d(d+1)}{2}\bar{\nabla}_{\alpha}\phi\bar{\nabla}^{\lambda}\bar{\nabla}^{\alpha}\phi-d(d-1)e^{-\phi}\bar{\nabla}^\lambda\phi\bigg) \bigg),\nonumber \\
 && \nabla^2R_{ij} = g_{ij}\bigg((d-1)e^{-\phi}(\bar{\nabla}^\alpha\phi\bar{\nabla}_\alpha\phi-\bar{\nabla}^2\phi)-\frac{1}{2}\bar{\nabla}^2(\bar{\nabla}^2\phi)-\frac{d}{2}(\bar{\nabla}_\mu\phi\bar{\nabla}^2\bar{\nabla}^\mu\phi\nonumber \\
 &&~~~~~~~~~~~~+\bar{\nabla}^\alpha\bar{\nabla}^\mu\phi\bar{\nabla}_\alpha\bar{\nabla}_\mu\phi)+\frac{\bar{\nabla}^\mu\phi}{2}(\bar{\nabla}^\nu\phi\bar{R}_{\mu\nu}-\frac{d}{2}\bar{\nabla}_\mu\bar{\nabla}^2\phi-\frac{d(d+1)}{2}\bar{\nabla}^\nu\phi\bar{\nabla}_\mu\bar{\nabla}_\nu\phi\nonumber \\
 &&~~~~~~~~~~~~-(d^2-1)e^{-\phi}\bar{\nabla}_\mu\phi+\frac{1}{2}\bar{\nabla}_\mu\phi\bar{\nabla}^2\phi) \bigg).\nonumber\\
\end{eqnarray}

\section{Effective equations in near horizon region coordinates}\label{rescale}
The ansatz metric has a fast direction $d\psi$ along which derivatives acting on the metric can be of order $\mathcal{O}(D)$. We find it convenient  to go to a ``scaled'' coordinate system based around any arbitrary point in the membrane region so that the maximum order of derivatives along the fast direction acting on the metric becomes  $\mathcal{O}(1)$. We use the procedure already mentioned in \cite{Bhattacharyya:2015dva, Bhattacharyya:2015fdk, Saha:2018elg} for this purpose. We briefly review the procedure in this section.


First of all we zoom into the membrane region about any arbitrary point $x_0^\mu$ using the following coordinate transformation,\footnote{We will use $\frac{1}{D-3}$, rather than $\frac{1}{D}$ as the perturbation parameter throughout this paper as this is the natural variable that appears in the blackening factor of black holes in arbitrary dimensions.}
\begin{eqnarray}
&&x^\mu=x^\mu_0+\frac{1}{D-3}~ \alpha^{\mu}_a y^a,\nonumber\\
&&y^a=(D-3)(x^\mu-x^\mu_0)\alpha^a_\mu.
\end{eqnarray}
Due to this coordinate change, the effective $p+3$ dimensional metric and the derivative of the dilaton field in the  `$y^a$' coordinates transform as
\begin{eqnarray}
&&g_{ab}=\frac{\partial x^\mu}{\partial y^a}\frac{\partial x^\nu}{\partial y^b}g_{\mu\nu}=\frac{1}{(D-3)^2}\alpha^\mu_a\alpha^\nu_bg_{\mu\nu},\nonumber\\
&&\partial_a\phi=\frac{1}{D-3}\alpha^\mu_a\partial_\mu\phi.
\end{eqnarray}
Since, coordinate lengths of the order of $\mathcal{O}(1)$ in $x^\mu$ coordinates is equivalent to coordinate lengths of the order of $\mathcal{O}(D)$ in $y^a$ coordinates, the maximum order of derivatives on the metric in the $y^a$ coordinates does indeed become $\mathcal{O}(1)$. But the metric now measures distances which are order $\mathcal{O}(1/D^2)$ as we cover coordinates distances which are order $\mathcal{O}(1)$ in the $y^a$ coordinates. To ameliorate this situation we scale up the metric and the field measuring the derivative of the dilaton field as

\begin{eqnarray}
 &&G_{ab}=(D-3)^2g_{ab}=\alpha^\mu_a\alpha^\nu_bg_{\mu\nu},\nonumber\\
&&G^{ab}=\frac{1}{(D-3)^2}g^{ab}=\alpha^a_\mu\alpha^b_\nu g^{\mu\nu},\nonumber\\
&&\chi_a=(D-3)\partial_a\phi=\alpha^\mu_a\partial_\mu\phi.
\end{eqnarray}
Here we have introduced a new field corresponding to the derivative of the dilaton field namely $\chi_a$. Doing the above-mentioned scalings of the coordinates and fields we now have a system where metrics measure $\mathcal{O}(1)$ distances for unit coordinate distances and also the maximum order of derivatives on the fields is $\mathcal{O}(1)$. We write down the  effective gravity equations obtained in the last section in these scaled system now. The equations in the $p+3$ directions of the reduced spacetime are given by
\begin{eqnarray}
\nonumber T_1-T_2-A-\frac{\tilde{\kappa}}{1-\epsilon}T_3+\frac{\tilde{\kappa}}{1-\epsilon}\big[2a_1 T_4-\big(2a_1+a_2+2a_3\big)T_5+\big(2a_1+\frac{a_2}{2}\big)T_6\\
-(2a_2+4a_3)T_7+(a_2+4a_3)T_8+2a_3 T_9-4a_3 T_{10}\big]=0,
\end{eqnarray}
where,
\begin{eqnarray}
T_1 &=&\bar{ R}_{ab}-\frac{d\epsilon}{2}\bar{\nabla}_a\chi_b
-\frac{d\epsilon^2}{4}\chi_a\chi_b,\nonumber\\
T_2 &=& \frac{G_{ab}}{2}\bigg(\bar{R}-d\epsilon \bar{\nabla}_a\chi^a -\frac{d(d+1)\epsilon^2}{4}\chi_a\chi^a+\frac{d(d-1)\epsilon^2}{\sigma^2}\bigg),\nonumber\\
T_3 &=& a_1 GB1+a_2 GB2+a_3 GB3,\nonumber\\
A &=&\frac{l}{2} G_{ab}(1+3\epsilon),\nonumber\\
GB1 &=& \frac{G_{ab} }{2}\bigg(\bar{R}-d\epsilon\bar{ \nabla}_c\chi^c -\frac{d(d+1)\epsilon^2}{4}\chi_c\chi^c+\frac{d(d-1)\epsilon^2}{\sigma^2}\bigg)^2 ,\nonumber\\
GB2 &=& \frac{G_{ab} }{2}\bigg[\big(\bar{R}_{cd}-\frac{d\epsilon}{2}\bar{\nabla}_c\chi_d-\frac{d\epsilon^2}{4}\chi_c\chi_d\big)\big(\bar{R}^{cd}-\frac{d\epsilon}{2}\bar{\nabla}^c\chi^d-\frac{d\epsilon^2}{4}\chi^c\chi^d\big)\nonumber\\ 
&&+ d\epsilon^4 \bigg(\frac{d-1}{\sigma^2}-\frac{1}{2\epsilon}\bar{\nabla}_c\chi^c
-\frac{d}{4}\chi_c\chi^c\bigg)^2\bigg],\nonumber\\
GB3 &=& \frac{G_{ab} }{2}\bigg[\bar{R}_{cdef}\bar{R}^{cdef}+d\epsilon^2\bigg(\bar{\nabla}_c\chi_d+\frac{\epsilon}{2}\chi_c\chi_d\bigg)\bigg(\bar{\nabla}^c\chi^d+\frac{\epsilon}{2}\chi^c\chi^d\bigg)\nonumber\\&&+\frac{2\epsilon^4}{\sigma^4}d(d-1) \bigg(\frac{\sigma^2}{4}\chi_c\chi^c-1\bigg)^2\bigg],\nonumber\\
T_4 &=&  \bigg(\bar{R}-d\epsilon \bar{\nabla}_c\chi^c -\frac{d(d+1)\epsilon^2}{4}\chi_c\chi^c+\frac{d(d-1)\epsilon^2}{\sigma^2}\bigg)\bigg(\bar{R}_{ab}-\frac{d\epsilon}{2}\bar{\nabla}_a\chi_b
-\frac{d\epsilon^2}{4}\chi_a\chi_b\bigg),\nonumber\\
T_5 &=& \bigg(\bar{\nabla}_a\bar{\nabla}_b\bar{R}-d\epsilon\bar{\nabla}_a\bar{\nabla}_b (\bar{\nabla}_c\chi^c)-\frac{d(d+1)\epsilon^2}{2}(\bar{\nabla}_a\chi_c\bar{\nabla}_b\chi^c+\chi^c\bar{\nabla}_a\bar{\nabla}_b\chi_c)\nonumber\\
 &&+\frac{d(d-1)\epsilon^3}{\sigma^2}(\epsilon\chi_a\chi_b-\bar{\nabla}_a\chi_b)\bigg),\nonumber\\
T_6 &=& \bigg(\bar{\nabla}^2\bar{R}-d\epsilon \bar{\nabla}^2(\bar{\nabla}_c\chi^c)-\frac{d(d+1)\epsilon^2}{2}(\chi^c\bar{\nabla}^2\chi_c+\bar{\nabla}_c\chi_d\bar{\nabla}^c\chi^d)\nonumber\\
&&+\frac{d(d-1)\epsilon^3}{\sigma^2}\big(\epsilon\chi_c\chi^c-\bar{ \nabla}_c\chi^c\big)+\frac{d}{2}\bigg(\epsilon\chi^c\bar{\nabla}_c\bar{R}-d\epsilon^2\chi^c\bar{\nabla}_c\bar{\nabla}_d\chi^d\nonumber\\
 &&-\epsilon^3\frac{d(d+1)}{2}\chi^c\chi^d\bar{\nabla}_c\chi_d 
-\frac{\epsilon^4d(d-1)}{\sigma^2}\chi^c\chi_c\bigg)\bigg)G_{ab},\nonumber
\end{eqnarray}
\begin{eqnarray}
T_{7}&=&\bigg[G^{gf}\bigg(\bar{R}_{eg}-\frac{d\epsilon}{2}\bar{\nabla}_e\chi_g
-\frac{d\epsilon^2}{4}\chi_e\chi_g\bigg){\bar{R}^e}_{abf}\nonumber\\
&& +d\epsilon^3\bigg(\frac{d-1}{\sigma^2}-\frac{1}{2\epsilon}\bar{\nabla}_c\chi^c
-\frac{d}{4}\chi_c\chi^c\bigg)\bigg(\frac{\bar{\nabla}_a\chi_b}{2}+\frac{\epsilon \chi_a\chi_b}{4}\bigg) \bigg],\nonumber\\
T_{8} &=&\bigg[\bar{\nabla}^2\bar{ R}_{ab}-\frac{d\epsilon}{2}\bar{\nabla}^2(\bar{\nabla}_a\chi_b)-\frac{d\epsilon^2}{2}\bar{\nabla}_c\chi_a \bar{\nabla}^c \chi_b-\frac{d\epsilon^2}{4}(\chi_a\bar{\nabla}^2\chi_b+\chi_b\bar{\nabla}^2\chi_a)+\frac{d}{2}\epsilon\chi_c\bar{\nabla}^c\bar{R}_{ab}
\nonumber\\ &&
-\frac{d}{4}\bigg(d\epsilon^2\chi^c\bar{\nabla}_c\bar{\nabla}_a\chi_b+\epsilon^2\chi^c(\chi_b\bar{R}_{ca}+\chi_a\bar{R}_{cb})\bigg)+\frac{d \epsilon^4}{2} \chi_a\chi_b\bigg(\frac{(d-1)}{\sigma^2}-\frac{\bar{\nabla}^c\chi_c}{2\epsilon}\bigg)\bigg], \nonumber\\
T_{9} &=& \bar{R}_{acde}{\bar{R}_b}^{cde}+ \frac{d\epsilon^2}{2} \bigg(\bar{\nabla}_a\chi_c+\frac{\epsilon}{2}\chi_a\chi_c\bigg)\bigg(\bar{\nabla}_b\chi^c+\frac{\epsilon}{2}\chi_b\chi^c\bigg),\nonumber\\
T_{10} &=& G^{cd} \bigg(\bar{R}_{ac}-\frac{d\epsilon}{2}\bar{\nabla}_a\chi_c
-\frac{d\epsilon^2}{4}\chi_a\chi_c\bigg)\bigg(\bar{R}_{bd}-\frac{d\epsilon}{2}\bar{\nabla}_b\chi_d
-\frac{d\epsilon^2}{4}\chi_b\chi_d\bigg).
\end{eqnarray}

The equations in the isometry directions which are proportional to the unit sphere metric $\Omega_{ij}$ are given by
\begin{eqnarray}
\nonumber t_1-t_2-\tilde{A}-\frac{\tilde{\kappa}}{1-\epsilon}t_3+\frac{\tilde{\kappa}}{1-\epsilon}\big[2a_1 t_4-\big(2a_1+a_2+2a_3\big)t_5+\big(2a_1+\frac{a_2}{2}\big)t_6\\
-(2a_2+4a_3)t_7+(a_2+4a_3)t_8+2a_3 t_9-4a_3 t_{10}\big]=0\nonumber,
\end{eqnarray}
where 
\begin{eqnarray*}
t_1 &=& \bigg(\frac{d-1}{\sigma^2}-\frac{1}{2\epsilon}\bar{\nabla}_a\chi^a-\frac{d}{4}\chi_a\chi^a\bigg),\nonumber\\
t_2 &=& \frac{1}{2}\bigg(\bar{R}\epsilon^{-2}-\frac{d}{\epsilon}\bar{\nabla}_a\chi^a-\frac{d(d+1)}{4}\chi_a\chi^a+\frac{d(d-1)}{\sigma^2}\bigg),\nonumber\\
\tilde{A}&=& \frac{l}{2}\epsilon^{-2},\nonumber\\
t_3 &=& \frac{1}{2} (a_1 gb1+a_2 gb2+a_3 gb3),\nonumber\\
gb1 &=& \epsilon^{-2}\bigg(\bar{R}-d\epsilon \bar{\nabla}_a\chi^a -\frac{d(d+1)\epsilon^2}{4}\chi_a\chi^a+\frac{d(d-1)\epsilon^2}{\sigma^2}\bigg)^2,\nonumber\\
gb2 &=& \epsilon^{-2} \big(\bar{R}_{ab}-\frac{d\epsilon}{2}\bar{\nabla}_a\chi_b-\frac{d\epsilon^2}{4}\chi_a\chi_b\big)\big({R}^{ab}-\frac{d\epsilon}{2}\bar{\nabla}^a\chi^b-\frac{d\epsilon^2}{4}\chi^a\chi^b\big)\\&&+d\bigg(\frac{d-1}{\sigma^2}\epsilon-\frac{1}{2}\bar{\nabla}_a\chi^a
-\frac{d\epsilon}{4}\chi_a\chi^a\bigg)^2,\nonumber\\
gb3 &=& \epsilon^{-2}\bar{R}_{abcd}\bar{R}^{abcd}+4d\bigg(\frac{\bar{\nabla}_a\chi_b}{2}+\frac{\epsilon}{4}\chi_a\chi_b\bigg)\bigg(\frac{\bar{\nabla}^a\chi^b}{2}+\frac{\epsilon}{4}\chi^a\chi^b\bigg)\nonumber\\&&+2\sigma^{-4}d(d-1)\epsilon^2\bigg(\frac{\sigma^2}{4}\chi_a\chi^a-1\bigg)^2,\nonumber\\
t_4&=& \bigg(\bar{R}-d\epsilon \bar{\nabla}_a\chi^a-\frac{d(d+1)\epsilon^2}{4} \chi_a\chi^a+\frac{d(d-1)\epsilon^2}{\sigma^2}\bigg)\bigg(\frac{d-1}{\sigma^2}-\frac{1}{2\epsilon}\bar{\nabla}_b\chi^b-\frac{d}{4}\chi_b\chi^b\bigg),\nonumber\\
t_5 &=& \frac{\epsilon^2}{2}\bigg(\frac{\chi^c\bar{\nabla}_c\bar{R}}{\epsilon^3}-\frac{d}{\epsilon^2}\chi^c\bar{\nabla}_c\bar{\nabla}_a\chi^a-\frac{d(d+1)}{2\epsilon}\chi^d\chi^c\bar{\nabla}_c\chi_d-\frac{d(d-1)}{\sigma^2}\chi_d\chi^d\bigg),\nonumber
\end{eqnarray*}
\begin{eqnarray}
t_6 &=& \epsilon^2\bigg[ \epsilon^{-4}\bar{\nabla}^2\bar{R}-d\epsilon^{-3}\bar{\nabla}^2\bar{\nabla}_a\chi^a-\frac{d(d+1)}{2\epsilon^2}\bigg(\chi_a\bar{\nabla}^2\chi^a+\bar{\nabla}^a\chi^b\bar{\nabla}_a\chi_b\bigg)\nonumber\\&&+\frac{d(d-1)}{\sigma^{2}}\big(\chi_a\chi^a-\frac{\bar{\nabla}_a\chi^a}{\epsilon}\big)+\frac{d}{2\epsilon^4}\bigg(\epsilon\chi^c\bar{\nabla}_c\bar{R}-d\epsilon^2\chi^c\bar{\nabla}_c\bar{\nabla}_d\chi^d\nonumber\\ &&-\epsilon^3\frac{d(d+1)}{2}\chi^d\chi^c\bar{\nabla}_c\chi_d 
-\epsilon^4\frac{d(d-1)}{\sigma^2}\chi_d\chi^d\bigg)\bigg],\nonumber\\
t_7 &=&  \epsilon^2 \bigg[\epsilon^{-2}\big(\bar{R}^{ab}-\frac{d\epsilon}{2}\bar{\nabla}^a\chi^b-\frac{d\epsilon^2}{4}\chi^a\chi^b\big)\bigg(\frac{\bar{\nabla}_a\chi_b}{2\epsilon}+\frac{\chi_a\chi_b}{4}\bigg)\nonumber\\
&& +\sigma^{-2}(d-1) \bigg(\frac{d-1}{\sigma^2}-\frac{1}{2\epsilon}\bar{\nabla}_a\chi^a-\frac{d}{4}\chi_a\chi^a\bigg)\bigg(\frac{\sigma^2}{4}\chi_a\chi^a-1\bigg)\bigg],\nonumber\\
t_8 &=& \epsilon^2\bigg[\frac{d-1}{\sigma^2}\bigg(\chi_a\chi^a-\frac{\bar{\nabla}_a\chi^a}{\epsilon}\bigg)-
\frac{1}{2\epsilon^3}\bar{\nabla}^2(\bar{\nabla}_a\chi^a)-\frac{d}{2\epsilon^2}\chi_a\bar{\nabla}^2\chi^a-\frac{d}{2\epsilon^2}\bar{\nabla}^a\chi^b\bar{\nabla}_a\chi_b   \nonumber \\ &&
+\frac{1}{2}\bigg(\frac{1}{\epsilon^2}\chi^c\chi^d\bar{R}_{cd}-\frac{d}{2\epsilon^2}\chi_c\bar{\nabla}^c\bar{\nabla}_d\chi^d-\frac{d(d+1)}{2\epsilon}\chi^c\chi^d\bar{\nabla}_c\chi_d-\frac{d^2-1}{\sigma^2}\chi^c\chi_c\nonumber\\&&+\frac{1}{2\epsilon}\chi_c\chi^c\bar{\nabla}_d\chi^d\bigg)\bigg],\nonumber\\
t_{9} &=& \epsilon^2 \bigg[2\bigg(\frac{\bar{\nabla}_a\chi_b}{2\epsilon}+\frac{1}{4}\chi_a\chi_b\bigg)\bigg(\frac{\bar{\nabla}^a\chi^b}{2\epsilon}
+\frac{1}{4}\chi^a\chi^b\bigg)+2\sigma^{-4}(d-1)\bigg(\frac{\sigma^2}{4}\chi_a\chi^a-1\bigg)^2\bigg],\nonumber\\
t_{10}&=& \epsilon^2  \bigg(\frac{d-1}{\sigma^2}-\frac{1}{2\epsilon}\bar{\nabla}_a\chi^a-\frac{d}{4}\chi_a\chi^a\bigg)^2 .
\end{eqnarray}

\section{A choice of coordinate system }\label{patch}
Having specified the set of equations that we will work with in the last section we will now  explain the choice of the coordinate system that we use for our computation in this section. To do that it will be useful to understand the reformulation of  the asymptotic metric of the $AdS$ black brane geometry which solves for the gravity equations \eqref{eom}
\begin{equation}
ds^2=-\tilde{l}r^2dt^2+\frac{dr^2}{\tilde{l}r^2}+l r^2dx_idx^i,\quad \text{where}\quad \tilde{l}=l+\tilde{\kappa}\tilde{\mathcal{A}}l^2,
\end{equation} 
 in terms of the effective spacetime. 

The metric written above has an $ISO(D-2)$ isometry. This metric can be recast in a manner where only a $SO(D-p-3)$ isometry is manifested in the following way
\begin{eqnarray}\label{split_poincare}
ds^2&=&-\tilde{l}r^2dt^2+\frac{dr^2}{\tilde{l}r^2}+l r^2(dy_idy^i+d\xi_id\xi^i),\nonumber\\
&=&-\tilde{l}r^2dt^2+\frac{dr^2}{\tilde{l}r^2}+l r^2(dy_idy^i+dS^2)+l r^2S^2d\Omega_{D-p-3}^2.
\end{eqnarray}
Here we have introduced an ad hoc spacial direction along $dS$ along which the manifested $ISO(D-2)$ isometry is broken into $SO(D-p-3)\times ISO(p)$. This way of writing the metric helps us to identify the effective metric and dilaton field for the spacetime to be given by
\begin{eqnarray}
ds^2_{eff}&=&-\tilde{l}r^2dt^2+\frac{dr^2}{\tilde{l}r^2}+l r^2(dy_idy^i+dS^2),\nonumber\\
\phi&=& 2\log(r S).
\end{eqnarray}
The shape and velocity functions in terms of which the ansatz metric is defined are functions of only the $p+3$ directions of the reduced spacetime\footnote{The choice of the number of directions `$p$' is completely arbitrary, and we will show later that all our results are invariant under this choice.}. In addition the velocity field has components only along these $p+3$ directions. Of the $p+3$ directions $3$ of the directions are distinct  namely those along: 1) The direction of the velocity field `$u$', 2) the direction of the normal to the membrane `$n\propto d\psi$' and 3) the direction along which the isometry is broken from $ISO(D-2)$ to $SO(D-p-3)$ i.e. `$dS$'. Before explicitly stating the choice of the coordinates let us evaluate the divergence of the velocity vector field in the coordinates \eqref{split_poincare}. It is given by
\begin{eqnarray}
\nabla\cdot u=\frac{D-2}{r}u^r+\frac{D-p-3}{S}u^S+\partial_{\mu}u^\mu.
\end{eqnarray}
We need to make sure that the divergence of the vector vanishes at leading order in large $D$ so that the leading order ansatz vanishes. We choose $u^r\sim\mathcal{O}(1/D)$ and $u^S\sim\mathcal{O}(1/D)$ for this purpose. This makes sure that the divergence of vector field is $\mathcal{O}(D^0)$ and we denote this as
\begin{equation}
\nabla\cdot u=\mathcal{U},
\end{equation}
where $\mathcal{U}$ is an $O(D^0)$ quantity. 

We now follow the convention of \cite{Bhattacharyya:2015dva, Bhattacharyya:2015fdk, Saha:2018elg} to choose $3$ of the coordinates to be along $n$, $O=n-u$,  and $dZ=dS-(n\cdot dS) dS$. We choose $O$ as one of the coordinate directions so that we have a null (ingoing) coordinate which makes it easier to impose regularity conditions at the horizon. The $dZ$ direction is chosen so that it is orthogonal to both $n$ and $u$ which is made sure by the choice of $u^r$ and $u^S$ mentioned above. We assign the following names to the coordinates in the membrane region about an arbitrary point with coordinates $x_0^\mu$
\begin{eqnarray}
&&R=(D-3)(\psi-1),\nonumber\\
&&V=(D-3) \left(x^\mu-x_0^\mu\right)O_\mu,\nonumber\\
&&z=(D-3)\left(x^\mu-x_0^\mu\right)Z_\mu,\nonumber\\
&&y^i=(D-3)\left(x^\mu-x_0^\mu\right)Y^i_\mu.
\end{eqnarray}
In terms of these coordinates the ansatz metric to leading order in large $D$ is given by
\begin{eqnarray}\label{ansatzincoord}
ds^2&=&2\frac{D-3}{\mathcal{K}}dVdR-\left(1-e^{-R}\right)\left(1-\tilde{\kappa}a_3\frac{\mathcal{K}^2}{(D-3)^2}e^{-R}\right)dV^2\nonumber \\
&&+\frac{l {r_0}^2}{1-l {r_0}^2 n_s^2}dz^2+l {r_0}^2 dy_idy^i,\nonumber\\
e^{\phi_0}&=&l r_0^2 S_0^2,
\end{eqnarray}
 where $\mathcal{K}= (D-3)\left(\sqrt{1-l r_0^2 n_s^2}\sqrt{l\left(1+\tilde{\kappa}\tilde{\mathcal{A}}l\right)}+\frac{n_s}{S_0}\right)$ with $n_s=n.dS$.

The ansatz metric is not a solution to the gravity equations at subleading orders in $1/D$. We need to add corrections to the ansatz metric and the dilaton field in a  $1/D$ expansion to get a solution to the gravity equations up to the desired order. The corrected metric and dilaton field can be parametrised as
\begin{eqnarray}\label{metric_expansion}
G_{\mu\nu}&=&g^{asym}_{\mu\nu}+\sum_{n=0}^\infty\left(\frac{1}{D-3}\right)^n\left(h_{\mu\nu}^{(n)}+\tilde{\kappa} h_{\mu\nu,\tilde{\kappa}}^{(n)}\right),\nonumber\\
\phi&=&\sum_{n=0}^\infty\left(\frac{1}{D-3}\right)^n\left(\phi^{(n)}+\tilde{\kappa}\phi^{(n)}_{\tilde{\kappa}}\right),
\end{eqnarray}
where, $$g^{asym}_{\mu\nu}+h_{\mu\nu}^{(0)}+\tilde{ \kappa}h^{(0)}_{\mu\nu,\tilde{ \kappa}},$$ is the effective metric part  and $\phi^{(0)}$ is the dilaton field part of the ansatz metric. 
 
 The gravity equations and their effective counterpart equations are a set of non-linear coupled partial differential equations and there are no well defined algorithmic procedure to extract analytic solutions out of  them.  The ansatz metric that we work with has a fast direction along $R$ direction in our coordinate system and hence, when the gravity equations act on the ansatz metric, the piece that is left unsolved at the subleading order is a function of the coordinate $R$ only with coefficients of the various $R$-dependent parts in the function being ultra-local functions of shape and velocity data. To be more precise, the derivatives transverse to the $R$ direction acting on the metric are always suppressed by $\mathcal{O}(1/D)$ w.r.t leading order in $R$ derivatives and hence, around a given point in the membrane region they only contain information about local data of the membrane like the extrinsic curvature of the membrane or the derivative of the velocity field about that point. These local data then appear as coefficient of the non-trivial $R$-dependent terms in the gravity equation at subleading order. For this reason the correction metric at this order can be thought of as function only of $R$ coordinate and the gravity equation acting on them produce homogeneous pieces in $R$ derivatives acting on the correction metric. Hence, the gravity equations that we need to solve become ordinary differential equations (ODEs) in $R$ coordinate acting on the correction pieces with sources being determined by local membrane data like extrinsic curvatures etc. In particular the differential equations that we arrive at are particularly simple to solve.  

 \section{Tensor structures of the equations}
 As explained above, in the large $D$ limit the gravity equations that we need to solve are ODEs involving the metric corrections with sources determined by the membrane data. 
 The corrections to the effective spacetime metric can be decomposed according to their tensor structures w.r.t the $p$ equivalent directions orthogonal to $u$, $n$ and $dZ$ directions. The ODEs can also be classified  according to this tensor structure in the sense that in a given tensor sector the homogeneous parts of the ODEs involve metric corrections of that particular sector only. Pushing the argument a bit further it is easy to see that the inhomogeneous parts of the ODEs comprising the shape and velocity functions also appear in the same tensor structure as the homogeneous parts. The different tensor sectors are completely decoupled. This is of particular use to us for solving the problem on a computer as each tensor structure can be solved separately. 
 
 Similar strategy has also been followed in \cite{Bhattacharyya:2015dva, Bhattacharyya:2015fdk,Dandekar:2016fvw,Saha:2018elg} where the independent shape and velocity data which can appear as sources for the 
 leading order ODEs have been classified in the tensor sectors in the following manner. 
 \begin{center}
 	\begin{table}[h]
 		\caption{Independent sources with tensor structures}\label{Rbasis}
 		\begin{tabular}{ |c|c|c| }
 			\hline
 			Scalar sector  & Vector sector & Tensor sector \\ 
 			\hline
 			$s_1 = u\cdot K\cdot u $& $ v_1^\mu=(u\cdot K)_\beta p^{\beta\mu} $ & $t_1^{\mu\nu}=p^{\mu\alpha}p^{\nu\beta}\left(K_{\alpha\beta}-\frac{p_{\alpha\beta}}{p}p^{\theta\phi}K_{\theta\phi}\right)$ \\ 
 			$s_2 = u\cdot K\cdot Z $& $v_2^\mu=z^\alpha K_{\alpha\beta}P^{\beta\mu}$& $t_2^{\mu\nu}=p^{\mu\alpha}p^{\nu\beta}\left(\nabla_{(\alpha}u_{\beta)}-\frac{p_{\alpha\beta}}{p}p^{\theta\phi}\nabla_{\theta}u_\phi\right)$ \\
 			$ s_3=Z\cdot K\cdot Z $ & $v_3^\mu=u^\alpha\nabla_\alpha u_\beta P^{\beta\mu}$     & $~$     \\
 			$s_4=P^{\mu\nu}K_{\mu\nu} $& $v_4^\mu=z^\alpha \nabla_\alpha u_\beta P^{\beta\mu}$ &  \\
 			$s_5=P^{\mu\nu}\nabla_\mu u_\nu $& $~$ & \\
 			\hline
 		\end{tabular}
 	\end{table}
 \end{center}
where,$$\quad p_{\mu\nu}=g_{\mu\nu}+u_\mu u_\nu-n_\mu n_\nu-\frac{l {r_0}^2z_{\mu}z_\nu}{1-l{r_0}^2n_S^2}.$$
 The above set of independent sources are obtained after imposing the orthonormality constraints 
 $$n\cdot n=1,\quad u\cdot u=-1\quad \text{and}\quad u\cdot n=0,$$ and the auxiliary conditions
 $$n\cdot\nabla  n=0\quad \text{and}\quad n\cdot\nabla u=0.$$

\section{The choice of gauge}
The metric corrections are arbitrary up to coordinate redefinitions (diffeomorphism). To get a well-defined corrected metric at subleading order, we need to make a choice of gauge so that the redundant diffeomorphism degrees of freedom are removed. We make the following gauge choice (found particularly useful in earlier works on the large $D$ membrane paradigm)
$$h_{\mu\nu}O^\mu=0.$$
This gauge choice when implemented in our choice of patch coordinates,  the most general form of metric correction in the different tensor sectors can be written as
\begin{eqnarray}\label{effective_space_metric}
&&h_{ab}^{(1)}dx^adx^b+\tilde{\kappa} h_{ab,\tilde{\kappa}}^{(1)}dx^adx^b=\nonumber\\&&(S_{VV}(R)+\tilde{\kappa} S_{VV,\tilde{\kappa}}(R))dV^2+2(S_{Vz}(R)+\tilde{\kappa} S_{Vz,\tilde{\kappa}}(R))dVdz
+(S_{zz}(R)+\tilde{\kappa} S_{zz,\tilde{\kappa}}(R))dz^2\nonumber\\&&(S_{tr}(R)+\tilde{\kappa} S_{tr,\tilde{\kappa}}(R))dy^idy^i+2(V_{Vi}(R)+\tilde{\kappa} V_{Vi,\tilde{\kappa}}(R))dVdy^i+2(V_{zi}(R)+\tilde{\kappa} V_{zi,\tilde{\kappa}}(R))dzdy^i\nonumber\\&&
+(T_{ij}(R)+ T_{ij,\tilde{\kappa}})dy^idy^j.
\end{eqnarray}

\section{Boundary conditions and regularity of solutions}
The equations of gravity that we work with \eqref{eom} and its counterpart in effective spacetime, have terms up to two derivatives acting on the Riemann tensor (hence, four derivatives acting on the metric). Naively it looks like that the ODEs that we need to solve in the large $D$ limit are of order four. But we work only to linear order in the variable $\tilde{ \kappa}$ and we work towards finding only those corrections to the metric that smoothly match to solutions of two derivative Einstein gravity in the $\tilde{ \kappa}\rightarrow 0$ limit. This is encoded in the way we have parametrised the corrections to the ansatz metric order-by-order in $\tilde{ \kappa}$ in the previous section. All of these together make sure that the ODEs we need to solve are of degree two for both $h^{(1)}_{ab}$ and $h^{(1)}_{ab,\tilde{ \kappa}}$. Since, the equations have order two, the solutions to them come with two undetermined constants. We explain below how to fix these constants. 

We impose the regularity conditions on the metric corrections that they are analytic functions everywhere in the membrane region. This we do since we work with a coordinate system where to start with there are no singularities anywhere in the membrane region. We also need to impose boundary condition that the metric corrections vanish outside  the membrane region, i.e. as $R\rightarrow\infty$. This makes sure that the metric of the spacetime approaches asymptotic $AdS$ spacetime exponentially fast. These boundary and regularity conditions fix most of the undetermined constants for us. 

The boundary and regularity conditions though are not enough to take care of all of the undetermined constants. We find that we are left with two undetermined constant in the scalar sector and a one vector number worth of constant in the vector sector of the metric corrections. The origin of this can be traced back to ambiguities in the definition of the shape and velocity functions in the ansatz metric as noticed in\cite{Dandekar:2016fvw,Bhattacharyya:2017hpj,Bhattacharyya:2018szu, Saha:2018elg,Kundu:2018dvx}. The ansatz metric that we use in this paper is of the form\footnote{This is also the form of the ansatz metric used in other membrane paradigm papers.}
\begin{equation}
g^{ans}_{\mu\nu}=\bar{g}_{\mu\nu}+F(\psi)O_\mu O_\nu.
\end{equation}
Here, the notation $F(\psi)$ means that $F$ is a function of spacetime coordinates via its dependence on the shape function $\psi$. The function $F(\psi)$ must satisfy the additional constraints 
\begin{eqnarray}
&&F(\psi=1)=1.
\end{eqnarray}
Under a generic small change in the shape and velocity function parametrised by $$\psi\rightarrow \psi+\delta \psi \quad \text{and}\quad \delta u\rightarrow u+\delta u,$$ the change in the ansatz metric is given by 
\begin{equation}
\delta g^{ans}_{\mu\nu}=\left(\frac{\delta F}{\delta \psi}\delta \psi O_\mu O_\nu -F \left(\delta u_\mu O_\nu+\delta u_\nu O_\mu\right)\right).
\end{equation}
The form of the function $F$\footnote{In all the examples of membrane paradigm till now the leading order derivative $F$ in the large $D$ limit comes from its dependence on $\psi^{D+q}$, where $q$ is a order one number.} is such that the leading order ansatz metric remains unaffected by changes in the shape and velocity function of the form 
\begin{equation}
u\rightarrow u+\frac{\delta \tilde{u}}{D} \quad \text{and}\quad \psi\rightarrow\psi+\frac{\delta\tilde{\psi}}{D^2},
\end{equation}
where, $\delta \tilde{u}$ and $\delta\tilde{\psi}$ are $\mathcal{O}(1)$. 
The redefinition of the velocity field is not entirely free but is constrained by the requirement that the redefinition keeps the velocity field unit normalised. This requires that
$$\delta \tilde{u}\cdot u=0.$$

Due to the above redefinition, the direction of the normal vector remains along $d\psi$ up to errors of order $\mathcal{O}(1/D^2)$. In absence of any redefinition, the surface $\psi=1$ is a null surface w.r.t the ansatz metric. After the redefinition, the norm of the normal  to the redefined surface $n\propto d\psi$ is given by 
\begin{eqnarray}
(d\psi\cdot d\psi)^{new}&&=d\psi\cdot d\psi\left(1-F(\psi)-\frac{\delta F}{\delta\psi}\delta \psi\right),\nonumber\\
&&=-d\psi.d\psi \frac{\delta F}{\delta\psi}\delta \psi\quad (\text{as} \quad F(\psi=1)=1).
\end{eqnarray}
Hence, under the change in the shape function the surface $\psi=1$ no longer remains null at the first subleading order in $1/D$. The redefinition also adds a subleading piece to the component of the metric proportional to $O_\mu O_\nu$ which does not vanish at $\psi=1$. Hence, the metric corrections that we get along $dV^2$ which do not vanish at the $R=0$ change the position of the horizon from $\psi=1$ at subleading order. In order to align the definition of the shape field so that the position of the horizon does not change at subleading order we need to impose the boundary conditions 
\begin{equation}
S_{VV}(R=0)=0\quad \text{and}\quad S_{VV,\tilde{\kappa}}(R=0)=0.
\end{equation}
This fixes one of the undetermined constants in the scalar sector.

Another interesting observation is that the velocity vector field raised w.r.t the asymptotic spacetime is the generator of the event horizon i.e.  the surface $\psi=1$ to leading order in $1/D$. Under the redefinition of the velocity field that we mention, it can be shown that the generator of event horizon on the surface $\psi=1$ is given by
\begin{equation}
n_\mu g^{\mu\nu}_{ansatz}=u^\nu-\frac{1}{D}\delta u^\nu.
\end{equation}
Also, the change in the velocity field induces a change in the metric component with one leg along $O$ vector and the other leg along directions orthogonal to $n$ and $u$. So, components of the metric correction $S_{VY},S_{VY,\tilde{ \kappa}}$ and $V_{Vi},V_{Vi,\tilde{ \kappa}}$ which are non-zero on the $R=0$ surface can induce change in the definition of velocity field so that it does not coincide with the generator of the event horizon at sub-leading order. In order to fix the definition of the velocity field so that it coincides with the generator of the event horizon we require the following boundary conditions 

\begin{equation}
S_{VY}(R=0)=0,S_{VY,\tilde{ \kappa}}(R=0)=0\quad \text{ and}\quad  V_{Vi}(R=0)=0,V_{Vi,\tilde{ \kappa}}(R=0)=0.
\end{equation}
This fixes the remaining undermined constants in the scalar and vector sectors. 

Imposing the boundary conditions that we mentioned here, in addition to the regularity and fall off conditions, the first order metric is completely determined in terms of the shape and velocity data of the dual membrane along with the auxiliary conditions. 

\section{The constraint Einstein equation and the membrane equations}
The second order differential equations that we need to solve acquire the properties of the Einstein-Hilbert equations. This is easy to see because the ODEs at linear order in $\tilde{ \kappa}$ have homogeneous parts containing the terms in the metric corrections given by $h^{(1)}_{ab,\tilde{ \kappa}}$. Since, these metric corrections already come with a $\tilde{ \kappa}$ multiplied with them their homogeneous parts are completely determined by the action of the two- derivative part of the Einstein equations on them. Hence, not all of the equations  are dynamical in nature\footnote{Our ODEs do not determine the time evolution of the system but instead can be thought of determining the ``dynamics'' along the $d\psi$ direction.}. Some of the equations are constraints on the metric data at any given slice of the evolution direction. These equations are given by
\begin{equation}\label{cons}
E_{\mu}=E_{\mu\nu}G^{\nu\alpha}n_{\alpha},
\end{equation}
where, $E_{\mu\nu}=0$ are the gravity equations, $G^{\nu\alpha}$ is the inverse of the metric of the spacetime and $n_{\alpha}$ is the normal to the $\psi=\text{constant}$ slices.

All other components of the Einstein equation are dynamical in nature in the sense that given the data on a given $\psi=\text{constant}$ slice, these equations determine the data at the subsequent slices. The interesting property of Einstein equations is that once we demand that the dynamical equations are solved everywhere and the constraint equations are solved  on any $\psi=\text{constant}$ slice then the constraint equations are solved everywhere in the spacetime. In the language of ODEs that we solve the dynamical equations are of order two and the constraint equations are of order one. 

 We separate the constraint equations according to their tensor structures into scalar and vector equations. Scalar constraints are obtained by picking the component of \eqref{cons} along $u$, $dz$ and $O$ directions. The scalar constraint equations of interest to us will be the components along $u$ and $dZ$. The structure of the equation along $u$ direction is given by
  \begin{eqnarray}
&& e^{-R} S_{Vz}+(e^{-R}-1) \frac{d S_{Vz}}{dR}=\mathcal{H}_1(R),\nonumber\\
&& \tilde{\kappa}\Bigg(e^{-R}S_{Vz,\tilde{\kappa}}+(e^{-R}-1) \frac{dS_{Vz,\tilde{\kappa}}}{dR}\Bigg)= \mathcal{H}_2(R),
\end{eqnarray}
where, $\mathcal{H}_1(R)$ and $\mathcal{H}_2(R)$ are functions of $R$ with membrane data dependent sources. Since, the constraint equations carry same information irrespective of the choice of slice, we evaluate the above equation at $R=0$. Using the boundary conditions on $S_{Vz}$ and $S_{Vz,\tilde{ \kappa}}$ and assuming that the metric corrections will evaluate to regular functions everywhere, it is clear that the homogeneous parts vanish at $R=0$. Hence, the above equations are satisfied at $R=0$ only if the sources of the homogeneous parts vanish at $R=0$. This is not a content less  statement as has been observed in \cite{Dandekar:2016fvw, Bhattacharyya:2017hpj, Bhattacharyya:2018szu, Saha:2018elg}. In fact  this condition puts the following constraint on the membrane data
 \begin{equation}\label{scalcon}
 \mathcal{U}=0.
 \end{equation}
 
The other scalar constraint equation along $dZ$ direction is given by
\begin{eqnarray}
&&e^{-R}\left(e^R-1\right) \frac{dS_{zz}}{dR}=\mathcal{G}_1(R),\nonumber\\
&&\tilde{\kappa} e^{-R}\left(e^R-1\right) \frac{dS_{zz,\tilde{\kappa}}}{dR}= \mathcal{G}_2(R).
\end{eqnarray}
The homogeneous parts of the above equation evaluate to zero at $R=0$ and following similar arguments as before, the above equation gives the following constraint on the membrane data
\begin{eqnarray}
&&l\frac{D-3}{\mathcal{K}}\mathcal{C}^2 {r_0}^2 \left(1-\tilde{\kappa}{a_3} \frac{\mathcal{K}^2}{(D-3)^2}  \right) s_3-2 s_2\mathcal{C}+\frac{1}{l {r_0}^2}\frac{\mathcal{K}}{D-3}\left(1+\tilde{\kappa}{a_3} \frac{\mathcal{K}^2}{(D-3)^2} \right)s_1\nonumber\\
&&-\frac{1-l \frac{\mathcal{K}^2}{(D-3)^2} r_0^2 S_0^2}{l^2{r_0}^4 {S_0}^2}+ \frac{\tilde{\kappa}a_3\Big(1+l{r_0}^2{S_0}^2\left(2l-\frac{\mathcal{K}^2}{(D-3)^2}\right)\Big)}{l^2  {r_0}^4 {S_0}^2}=0,
\end{eqnarray}
where $\mathcal{C}= \bigg( \frac{1}{l{r_0}^2S_0}-\frac{n_s \sqrt{l(1+a_1\tilde{\kappa}l)}}{  \sqrt{1-l{r_0}^2{n_s}^2}}\bigg).$
  The vector component of the constraint equation evaluates to 
\begin{eqnarray}
&&e^{-R}\left(e^R-1\right)f_1 \frac{dV_{zi}}{dR} = \mathcal{C}_i(R),\nonumber\\
&&\tilde{\kappa} e^{-R}\left(e^R-1\right)\frac{dV_{zi,\tilde{\kappa}}}{dR}= \mathcal{C}_{i,\tilde{\kappa}}(R).
\end{eqnarray}
The homogeneous part of the above equation again vanishes at $R=0$ and hence, puts the following constraint on the vector data of the membrane

   \begin{equation}\label{veccon}
\bigg(l {r_0}^2 \frac{D-3}{\mathcal{K}}\mathcal{C} \Big(v_4^\mu-\Big(1-\tilde{\kappa} a_3\Big(\frac{\mathcal{K}}{(D-3)}\Big)^2\Big)v_2^\mu\Big)+v_1^\mu-\Big(1+\tilde{\kappa} a_3 \Big(\frac{\mathcal{K}}{(D-3)}\Big)^2\Big)v_3^\mu\bigg)p_{\mu i}=0.
\end{equation}

The constraint on the membrane data obtained are in fact dynamical equations on the membrane data and hence we call them the ``membrane equations". Our procedure to correct the ansatz metric to arrive at solutions to the gravity equations at subleading order in $1/D$  involves ``patching'' together many local solutions to form a global solution. The membrane equations constrain one-derivative data of the membrane and hence can be thought of as the leading order obstruction to patching together any arbitrary data to form a black hole solution. If and only if the membrane satisfies the membrane equations written above, one can construct a first order corrected solution from the membrane data. 
\section{Solution to the dynamical equations}\label{sol}
Once the membrane equations are satisfied, the dynamical equations can be solved to arrive at regular corrections to the ansatz metric at first order in $1/D$ by demanding regularity and imposing the boundary conditions on the metric mentioned earlier. The metric correction in the different tensor sectors obtained are given by

\subsection{Scalar Sector}

\begin{eqnarray}
S_{VV}&=& e^{-R} R \left(2-\frac{R(D-3)^2 \left(1-l^2 \left(1-l{r_0}^2{n_s}^2\right) {\mathcal{C}}^2 {r_0}^4 {S_0}^2\right)}{l {\mathcal{K}}^2 {r_0}^2 {S_0}^2}\right)+\frac{e^{-R}R(2+R)(D-3)}{2\mathcal{K}}s_1\nonumber\\
&&-\frac{e^{-R}R^2 l {r_0}^2(D-3)^2 \mathcal{C}}{\mathcal{K}^2}s_2+\tilde{\kappa}\Bigg( \frac{ {a_3} {\mathcal{K}} e^{-2 R}}{2(D-3)} \left(e^R (R (3 R+2)+4)-2 \left(R^2+2\right)\right)s_1+\nonumber\\
&&2 {a_3} l {r_0}^2\mathcal{C} e^{-2 R} \left(1-e^R\right) (R-1) R  s_2+\frac{{a_3} e^{-2 R} R}{l {r_0}^2 {S_0}^2}\Big(l^2 \left(e^R (R+2)-2\right) {r_0}^2 {S_0}^2\nonumber\\
&&-\left(e^R (R-4)-2 R+4\right)\left(1-l^2 \left(1-l{r_0}^2{n_s}^2\right) {\mathcal{C}}^2  {r_0}^4 {S_0}^2\right)\Big)\Bigg),\nonumber
\end{eqnarray}
\begin{eqnarray}
S_{VZ}&=& -\frac{l (D-3)e^{-R} R}{\left(1-l{r_0}^2{n_s}^2\right) \mathcal{C}\mathcal{K}}\left(1+\frac{\mathcal{K}}{l(D-3)}s_1-{r_0}^2\mathcal{C} s_2\right)+\tilde{\kappa}\Bigg(\frac{{a_3} l \mathcal{K}e^{-2 R} \left(1-3 e^R\right) R}{\left(1-l{r_0}^2{n_s}^2\right) (D-3)\mathcal{C}}\Big(1\nonumber\\
&&+\frac{\mathcal{K}}{l(D-3)}s_1-{r_0}^2\mathcal{C}s_2\Big)+a_1\left(-\frac{l^3(D-3) e^{-R} R {r_0}^2 {S_0}}{{\mathcal{K}} (1-\frac{l  {r_0}^2{n_s} \mathcal{K} {S_0}}{D-3})}\right)\Bigg),\nonumber\\
S_{zz,\tilde{\kappa}} &=& \frac{2a_3 l e^{-R}{\mathcal{K}}^2}{{\left(1-l {r_0}^2{n_s}^2\right)}^2(D-3)^2{\mathcal{C}}^2}\left(1+\frac{\mathcal{K}}{l(D-3)}s_1-{r_0}^2\mathcal{C} s_2\right),\nonumber\\
S_{tr,\tilde{\kappa}} &=& a_3 e^{-R}l{r_0}^2 \frac{\mathcal{K}}{D-3}\left(2\frac{\mathcal{K}}{D-3}+\left(s_5-s_4\right)\right),\nonumber\\
\delta \phi^{2} &=&- \frac{1}{l {r_0}^2}\left(2 S_{tr}+(1-l{r_0}^2{S_0}^2 )S_{zz}\right),
\end{eqnarray}
\subsection{Vector Sector}
\begin{eqnarray}
\nonumber V_{zi} &=& \tilde{\kappa}a_3 \frac{e^{-R}}{(1-l {r_0}^2 {n_s}^2)}\frac{l^2 {r_0}^4 \mathcal{K}}{D-3}\bigg(\frac{S \sqrt{1-l {r_0}^2 {n_s}^2}\mathcal{K}}{(D-3)(\sqrt{1-l {r_0}^2 { n_s}^2}-l^{3/2} {r_0}^2 S n_s)}v_{3\mu}-v_{2\mu}\bigg)p^\mu_{i} ,\\
\nonumber V_{Vi} &=& l {r_0}^2 R e^{-R}  \frac{D-3}{\mathcal{K}} (v_{1\mu}-v_{3\mu})p^\mu_{i}\\
\nonumber && +\tilde{\kappa} a_3 l {r_0}^2 R e^{-2R}\Big(l {r_0}^2 e^R \mathcal{C} v_{2\mu}+ \frac{\mathcal{K}}{D-3}(e^R-1)v_{1\mu}+\frac{\mathcal{K}}{D-3}(1-2e^R)v_{3\mu}\Big)p^\mu_{i} ,
\end{eqnarray}
where $\mathcal{C}= \bigg( \frac{1}{l{r_0}^2S_0}-\frac{n_s \sqrt{l(1+a_1\tilde{\kappa}l)}}{  \sqrt{1-l{r_0}^2{n_s}^2}}\bigg),$
\subsection{Tensor Sector}

\begin{eqnarray}
T_{\mu\nu}&=&-4\tilde{\kappa} a_3 e^{-R}\left(t_{1\mu\nu}-t_{2\mu\nu}\right)\frac{\mathcal{K}}{D-3},
\end{eqnarray}
where $\quad p_{\mu\nu}=g_{\mu\nu}+u_\mu u_\nu-n_\mu n_\nu-\frac{l {r_0}^2z_{\mu}z_\nu}{1-l{r_0}^2n_S^2}\quad \text{and}\quad n_S=n.dS.$

\section{Geometrized metric correction}\label{geo met}
All the metric corrections we have written down heavily use the presence of the $dS$ direction. The $dS$ direction is specific to the choice of what the value of  $p$ is. It has been beautifully explained in \cite{Bhattacharyya:2015fdk} that a solution which preserves a $SO(D-p-3)$ isometry also preserves a $SO(D-q-3)$ isometry for $q\gg p$. This is possible only if the solutions obtained by above secretly possess a symmetry under redistribution of finite number of coordinates between isometry and non-isometry directions. The mechanism by which the solutions possess this symmetry has been explained in \cite{Bhattacharyya:2015fdk}. The idea is that the solution obtained can be written in terms of geometric quantities of a membrane propagating in full spacetime (not just the effective $p+3$ dimensional spacetime) so that the dynamics preserves $SO(D-p-3)$ isometry for any arbitrary $p$. This mechanism was called ``geometrisation'' of the solutions. 

The full spacetime has only two distinct directions: a) along the  velocity vector field and b) along the normal to the membrane. The metric corrections that need to be added to the ansatz metric can be decomposed into tensor structures with respect to the directions orthogonal to these two distinct directions. Together with our gauge choice, i.e $H_{MN}O^N=0$, the most general metric correction in the tensor decomposition mentioned above can be written as

\begin{eqnarray}\label{geom_metric}
ds^2=ds_{0}^2+\frac{1}{D}\left(H_s(O_A dx^A)^2+H_A^{(V)}O_B dx^A dx^B+H_{AB}^{(T)}dx^A dx^B+\frac{1}{D}H^{Tr}P_{AB}dx^A dx^B\right),\nonumber\\
\end{eqnarray}
where $P_{AB}$ is the projector orthogonal to $n$ and $u$ vectors and
\begin{eqnarray}\label{geometric_met}
H_A^{(V)}n^A=0\quad, H_A^{(V)}u^A=0,\quad H^{(T)}_{AB}P^{AB}=0,\quad H^{(T)}_{AB}n^A=0,\quad h^{(T)}_{AB}u^A=0.
\end{eqnarray}

We need to put an extra factor of $1/D$ in front of the $H^{Tr}$  to make sure that the corrections  do not destroy the property that the leading order ansatz metric solves for the gravity equations at leading order in $1/D$. 

The map between the metric correction written in this manner and the metric correction written in the effective spacetime formalism is explicitly given in Appendix \eqref{geo_to_eff_map}. This can be used to write the expressions of the metric correction in geometric form as given here
\begin{eqnarray}
H_s&=&\left(2+\left(\frac{l (D-3)}{\mathcal{K}}-1\right)R\right)R e^{-R}-\frac{D-3}{\mathcal{K}}\frac{u.\nabla \mathcal{K}}{\mathcal{K}}R^2 e^{-R}+\frac{D-3}{\mathcal{K}} u.K.u \bigg(1+\frac{R}{2}\bigg)R e^{-R}\nonumber\\
&&+\tilde{\kappa} a_1 \frac{l^2(D-3)^2}{\mathcal{K}^2}R^2 e^{-R}+\tilde{\kappa} a_3 R\left(\frac{\mathcal{K}^2}{(D-3)^2}\left(4-R\right) +2 l(R-1)\right)e^{-R} \nonumber\\&& +2\tilde{\kappa} a_3 R\left( \frac{\mathcal{K}^2}{(D-3)^2}(R-2)-l(R-1)\right)e^{-2R}-2\tilde{\kappa} a_3 \frac{u.\nabla \mathcal{K}}{D-3} R(R-1)(e^R-1)e^{-2R}\nonumber\\
&& -\tilde{\kappa} a_3 \frac{\mathcal{K}}{2(D-3)}u.K.u\left(2(2+R^2)e^{-2R}-(4+2R+3R^2)e^{-R}\right),\nonumber\\
 H^{(V)}_M&=&\frac{R~e^{-R}(D-3)}{\mathcal{K}}\left(u^B K_{BA}-u.\nabla u_{A}\right)P^A_{M}\nonumber\\
&&+\tilde{\kappa}a_3 R~e^{-2R}\left(\frac{\nabla_B\mathcal{K}}{D-3} e^R+\frac{\mathcal{K}}{D-3}u^{A}K_{AB}(e^R-1)+\frac{\mathcal{K}}{D-3}(1-2e^R)u.\nabla u_B\right)P^{B}_{M},\nonumber\\
H^{(T)}_{MN}&=&-4\tilde{\kappa} a_3 e^{-R}\frac{\mathcal{K}}{D-3}\mathcal{T}_{MN},\nonumber\\
H^{Tr}&=&\mathcal{O}\left(\frac{1}{D}\right),
\end{eqnarray}
\begin{eqnarray}
\text{where,}\quad\mathcal{T}_{MN}&=&P_M^A P_{N}^{B}\left(K_{AB}-U_{AB}\right)-\frac{P_{MN}}{D-3}P^{TF}\left(K_{TF}-U_{TF}\right),\nonumber\\
\text{and}\quad U_{AB}&=&\frac{\nabla_A u_B+\nabla_B u_A}{2},\nonumber\\
\text{with}\quad  R&=&D(\psi-1)\quad \text{so that}\quad e^{-R}=\psi^{-D}.
\end{eqnarray}
\section{Geometrized membrane equation}\label{geo mem}
The equations determining the ``membrane equations'' can also be written in a geometric way. In the geometric way of writing, the ``vector'' membrane equations can be written in terms of the geometric vector data along the direction orthogonal to the $u$ and $n$ vectors as 
\begin{eqnarray}\label{veceq}
&&\bigg(\frac{\nabla^2 u_M}{\mathcal{K}}+u^N K_{MN}-\frac{\nabla_M  \mathcal{K}}{\mathcal{K}}\bigg(1- \frac{\tilde{\kappa}a_3\mathcal{K}^2}{(D-3)^2}\bigg)-(u.\nabla) u_M\bigg(1+\frac{\tilde{\kappa}a_3\mathcal{K}^2}{(D-3)^2}\bigg)\bigg)\mathcal{P}^M_A\nonumber\\
&&=O(1/D),
	\end{eqnarray}
where $\mathcal{P}^M_N=\delta^M_N+u^Mu_N$ is the projector orthogonal to the velocity vector in the membrane world-volume. In the above equation $K_{MN}$ is the extrinsic curvature of the membrane surface and $\mathcal{K}$ is the trace of the extrinsic curvature of the membrane. The covariant derivative and the dot products in the above equation are taken w.r.t asymptotic $AdS$ spacetime of the black hole with effective radius $L_{AdS}^2=\frac{1}{l(1+\tilde{ \kappa}a_1l)}$. 
The interpretation of the constraint on scalar membrane data coming from the constraint gravity equation projected along the velocity vector is already manifested in \eqref{scalcon} and is given by
\begin{equation}
\nabla.u=0.
\end{equation}
Giving geometric interpretation to the other scalar membrane equation requires a bit more work. For this we need to calculate the divergence of the  geometric vector membrane equation written above to leading order in $1/D$. This is given by
\begin{eqnarray}
\bigg(\frac{\nabla^M\nabla^2 u_M}{\mathcal{K}}+u\cdot \nabla K-\frac{\nabla^2  \mathcal{K}}{\mathcal{K}}\bigg(1- \frac{\tilde{\kappa}a_3\mathcal{K}^2}{(D-3)^2}\bigg)-\nabla^M(u.\nabla u_M)\bigg(1+\frac{\tilde{\kappa}a_3\mathcal{K}^2}{(D-3)^2}\bigg)\bigg)=0.\nonumber\\
\end{eqnarray}
It can be shown easily that to leading order in large $D$
\begin{eqnarray}\label{divrel}
&&\nabla^M\nabla^2u_M=\nabla^C(R_{DC}u^D),\nonumber\\
\text{and,}\quad &&\nabla^M(u\cdot \nabla u_M)=u^C R_{CD} u^D,
\end{eqnarray}
where, $R_{AB}$ is the Ricci tensor of the induced metric on the world-volume of the membrane.  Let the intrinsic curvature of the background spacetime be given by $\mathcal{R}_{\alpha\beta\gamma\delta}, \mathcal{R}_{\alpha\beta},\mathcal{R}$. The relation between the intrinsic curvature of the induced metric on the membrane and the extrinsic curvature of the membrane are given by the Gauss-Codacci relations 
\begin{eqnarray}\label{gauss_codacci}
\nabla^C K_{BC}-\nabla_B\mathcal{K}=\mathcal{R}_{\mu\beta}n^\mu e^\beta_B,\nonumber\\
\mathcal{R}_{\alpha \beta \gamma\delta}e^\alpha_A e^\beta _B e^\gamma _C e^\delta _D=R_{ABCD}+\left(K_{AD} K_{BC}-K_{AC} K_{BD}\right).
\end{eqnarray}
Using \eqref{divrel} and \eqref{gauss_codacci} in the divergence of the vector membrane equation one gets
\begin{equation}\label{divofeq}
\left(1-\tilde{\kappa}a_3\frac{\mathcal{K}^2}{(D-3)^2}\right)\frac{\nabla^2\mathcal{K}}{\mathcal{K}^2}+\left(1+\tilde{\kappa}a_3 \frac{\mathcal{K}^2}{(D-3)^2}\right)\Big(u.K.u+\frac{u\cdot \mathcal{R}\cdot u}{\mathcal{K}} \Big)-2\frac{u.\nabla \mathcal{K}}{\mathcal{K}}=\mathcal{O}\left(\frac{1}{D}\right).
\end{equation}
It can easily be checked that the above equation when written in terms of the data of the membrane written in effective spacetime language matches with the second scalar constraint on the membrane data coming from the constraint gravity equation evaluated along $dZ$. 
\subsection{Comparison to Charged Membranes with cosmological constant}
The leading order membrane equations  dual to charged black hole dynamics in presence of a cosmological constant were derived in \cite{Kundu:2018dvx} and are given by
\begin{equation}\label{membraneeqnts}
\begin{split}
\nabla\cdot u={\cal O}\left( \frac{1}{D}\right)\\
\frac{\nabla^2 Q}{\mathcal{K}} - u.\nabla Q - Q\left[ \frac{u.\nabla\mathcal{K}}{\mathcal{K}} - u\cdot K\cdot.u  - \frac{u\cdot\mathcal{R}\cdot u}{\mathcal{K}}\right] ={\cal O}\left( \frac{1}{D}\right)\\
\Big [ \frac{ \nabla^2 u_\nu}{\mathcal{K}}-(1-Q^2)\frac{\nabla_\nu \mathcal{K}}{\mathcal{K}}+u^\alpha K_{\alpha\nu}-(1+Q^2)(u.\nabla u_\nu)\Big]\mathcal{P}^\nu_\mu={\cal O}\left( \frac{1}{D}\right)
\end{split}
\end{equation}
It is particularly interesting to observe that when in the first and last equation written above we use the map $$Q\rightarrow\sqrt{\tilde{ \kappa}a_3}\frac{\mathcal{K}}{D}$$ we arrive at the geometric form of the scalar and vector membrane equations derived by us.
Also under this map the equation determining the charge variable reduce to
\begin{eqnarray}
\sqrt{\tilde{ \kappa}a_3}\left(\frac{\nabla^2 \mathcal{K}}{\mathcal{K}} - 2u.\nabla \mathcal{K} + \mathcal{K}\left[ u\cdot K\cdot.u  + \frac{u\cdot\mathcal{R}\cdot u}{\mathcal{K}}\right]\right) ={\cal O}\left( \frac{1}{D}\right)
\end{eqnarray}
which to the relevant order is the divergence of vector equation written above. The map between the charge and shape variable mentioned above is precisely which converts the leading order in $1/D$ ansatz metric written in \cite{Kundu:2018dvx} to the ansatz metric written by us in \eqref{ansatz}. We find this map\footnote{Also, observed for asymptotic flat spacetime in \cite{Saha:2018elg}.} very intriguing and we think it may be useful to understand its origin better. 

Another interesting thing we notice in this context is that naively the charge equation written above seems to have a piece which is not covariantisation of the flat space answer obtained in \cite{Bhattacharyya:2015fdk}, namely the piece $\frac{u\cdot\mathcal{R}\cdot u}{\mathcal{K}}$. But it is worth noticing that for both the charge equation and the divergence of the vector equation mentioned above this factor always comes int he combination $$u\cdot K\cdot.u  + \frac{u\cdot\mathcal{R}\cdot u}{\mathcal{K}}$$
which can also be written as $\frac{\nabla^M(u\cdot\nabla u_M)}{\mathcal{K}}$. Written this way this piece is still a covariantisation of the answer written in flat spacetime. We think that this is a more natural way to express the membrane equations from the context of the expression of the charge current derived in \cite{Bhattacharyya:2016nhn}, the relevant part of which is given by
 \begin{equation}
 J^\mu\propto\left(\mathcal{K} u^\mu-\frac{p^{\nu\mu}\nabla_\nu Q}{Q}-(u\cdot \nabla)u^\mu-\frac{\nabla^2 u^\mu}{\mathcal{K}}+(u\cdot K)^\mu\right)
 \end{equation}
 The conservation of the above charge current gives the charge membrane equation and in the conservation equation the term that appears is $\nabla^\mu(u\cdot \nabla)u_\mu$. So, in summary we can conclude that all the membrane equations in asymptotic $AdS$ spacetime that we know of at leading order in large $D$ are covariantisation of the answer in asymptotic flat spacetime. 
\section{Stress tensor}
The membrane dual to black holes in leading order in large $D$ has a world-volume stress tensor in the sense that its conservation gives rise to the membrane equations. This stress-tensor is given by
\begin{eqnarray}
T_{MN}&=&\frac{\mathcal{K}}{2}\left(1+\frac{\tilde{\kappa} a_3\mathcal{K}^2}{D^2}\right)u_M n_N+\left(1-\frac{\tilde{\kappa} a_3\mathcal{K}^2}{D^2}\right)\frac{K_{MN}}{2}-\frac{\nabla_M u_N+\nabla_N u_M}{2} \nonumber\\&&-\left(u_M V_N+u_N V_M\right), \nonumber\\
\text{where,}&&\nonumber\\
\quad V_M&=& -\frac{1}{2}\left(1-\frac{\tilde{\kappa} a_3 \mathcal{K}^2}{D^2}\right)\frac{\nabla_M\mathcal{K}}{\mathcal{K}}+\frac{\tilde{\kappa} a_3 \mathcal{K}^2}{D^2}(u.K)_M-\frac{\tilde{\kappa} a_3 \mathcal{K}^2}{2D^2} u\cdot \nabla u_M\nonumber\\&&+\left(1+\frac{\tilde{\kappa} a_3 \mathcal{K}^2}{D^2}\right)\frac{\nabla^2 u_M}{\mathcal{K}}.
\end{eqnarray}
The equation for  conservation of the stress tensor along $u_m$ is given by
\begin{eqnarray}
&&\nabla^M T_{MN}u^N=0\nonumber\\
\implies &&-u\cdot \nabla \left(\frac{\mathcal{K}}{2}\left(1+\frac{\tilde{\kappa} a_3\mathcal{K}^2}{D^2}\right)\right)-\frac{\mathcal{K}}{2}\left(1+\frac{\tilde{\kappa} a_3\mathcal{K}^2}{D^2}\right)\nabla\cdot u+\nabla\cdot V\nonumber\\
&&+\left(1-\frac{\tilde{\kappa} a_3\mathcal{K}^2}{D^2}\right)\nabla^M K_{MN}u^N-\frac{\nabla^2 u_Nu^N}{2}-\frac{\nabla^M \nabla_N u_M u^N}{2}=0.
\end{eqnarray}
Using the Gauss-Codacci relations given in the last section one can arrive at the following useful identities
\begin{equation}
\nabla^M\nabla_N u_M u^N=\nabla^M(u\cdot \nabla u_M)+\mathcal{O}(1)=\mathcal{K} u\cdot K\cdot u-\frac{\mathcal{R}}{D}+\mathcal{O}(1) ,
\end{equation}
\begin{equation}
\nabla^M\nabla^2 u_M=\nabla^A R_{BA} u^B=\mathcal{K}u \cdot \nabla \mathcal{K}.
\end{equation}
Combining the above equations we get
\begin{eqnarray}
\frac{\mathcal{K}}{2}\left(1+\frac{\tilde{\kappa} a_3\mathcal{K}^2}{D^2}\right)\nabla\cdot u&=&u\cdot \nabla\mathcal{K}-\frac{1}{2}\left(1-\frac{\tilde{\kappa} a_3 \mathcal{K}^2}{D^2}\right)\frac{\nabla^2\mathcal{K}}{K}-\nonumber\\&&\frac{\mathcal{K}u\cdot K\cdot u-\frac{\mathcal{R}}{D}}{2}\left(1+\frac{\tilde{\kappa} a_3\mathcal{K}^2}{D^2}\right).\nonumber\\
\end{eqnarray}
The right hand side is precisely the divergence of the vector membrane equation of motion and is $\mathcal{O}(1)$ and hence, we arrive at the scalar membrane equation
\begin{equation}
\nabla\cdot u=\mathcal{O}(1/D).
\end{equation}
Similarly, the conservation of the stress tensor orthogonal to the velocity vector in the membrane world-volume gives
\begin{eqnarray}
\left(\frac{\mathcal{K}}{2}\left(1+\frac{\tilde{\kappa} a_3 \mathcal{K}^2}{D^2}\right) u\cdot  \nabla u_N+\left(1-\frac{\tilde{\kappa} a_3\mathcal{K}^2}{D^2}\right)\frac{\nabla_N\mathcal{K}}{2}-\frac{\nabla^2 u_N}{2}-\nabla^M\nabla_N u_M\right)\mathcal{P}^N_A=0,\nonumber\\
\end{eqnarray}
which upon using the Gauss-Codacci relation gives precisely the vector membrane equation of motion. Hence, the membrane dynamics can be stated in a very compact form as the conservation of the stated stress energy tensor in the world-volume of the membrane. 
\section{Light quasi-normal modes}\label{qnm}
It has been shown in \cite{Bhattacharyya:2015dva, Bhattacharyya:2015fdk, Dandekar:2016fvw, Bhattacharyya:2017hpj, Bhattacharyya:2018szu} that the large $D$ membrane equations can predict the spectrum of light quasi-normal modes. The membrane equations can be used to predict the spectrum of light modes for situations where an explicit answer is not known from gravity side. In \cite{Chen:2017hwm} the spectrum of light quasi-normal modes  for static black holes in Gauss-Bonnet gravity was predicted from the effective equations derived in terms of effective mass and momentum variables. The prediction for these modes from the effective membrane equations in \cite{Saha:2018elg} matched with the results derived in \cite{Chen:2017hwm} in the common regime of validity. In this section we compute the spectrum of linearised fluctuations about a static spherical membrane and a static planar membrane to predict the spectrum of light quasi-normal modes for static black holes and static black branes in the theory of gravity under consideration. We start with the computation for spherical membrane in the next subsection. 
\subsection{Light quasi-normal modes of black hole}
To study the spectrum of linearised fluctuations about static-round membranes we consider a coordinate system adapted to the spherical symmetry of the problem. The metric of the $AdS$ spacetime to which the black holes asymptote to and in which the membrane propagates, in these coordinates is given by
\begin{equation}\label{abh}
ds^2=-dt^2\left(1+l r^2+\tilde{\kappa} \tilde{\mathcal{A}} l^2 r^2\right)+\frac{dr^2}{1+l r^2+\tilde{\kappa} \tilde{\mathcal{A}} l^2 r^2}+r^2d\Omega_{D-2}^2.
\end{equation}
We will express all our answers in terms of the variable $\tilde{l}$ given by
\begin{equation}
l(1+\tilde{\kappa} \tilde{\mathcal{A}} l)=\tilde{l}=\frac{1}{L_{AdS}^2},
\end{equation}
where, the effective $AdS$ length scale of the $AdS$ spacetime is given by $L_{AdS}^2$. 
Let the static spherical surface be given by $$r=1.$$ We will be concerned with small fluctuations about this configuration which can be parametrised as
\begin{equation}
r=1+\epsilon \delta r(t,\theta^a).
\end{equation}
The unit normalised space-like normal to this surface is given by
\begin{equation}
n=\frac{dr-\epsilon \partial_t\delta r-\epsilon \partial_a\delta r}{\sqrt{1+\tilde{l}r^2}}.
\end{equation}
The extrinsic curvature is given by
$$K^s_{AB}=\frac{\nabla_A n_B+\nabla_B n_A}{2}-\frac{n\cdot \nabla n_A n_B+n\cdot \nabla n_B n_A}{2}.$$

The superscript $s$ indicates the fact that this is the extrinsic curvature computed as a spacetime quantity. In the membrane equation of motion the pull back of the extrinsic curvature on to the membrane world-volume enters and the relation between the two is given by
\begin{equation}
K_{\mu\nu}=K_{AB}^s\frac{\partial x^A}{\partial x^\mu}\frac{\partial x^B}{\partial x^\nu},
\end{equation}
where $x^A$ are the spacetime coordinates and $x^\mu$ are the world-volume coordinates. The above equation is to be evaluated at the position of the membrane given by$$r=1+\epsilon \delta r.$$ The non-zero components of the pulled back extrinsic curvature are given by\footnote{The details of the computation are provided in the appendix \eqref{details_qnm_Bhole}.}
\begin{eqnarray}
&&K_{tt}=-\tilde{l}\sqrt{1+\tilde{l}}-\frac{\epsilon \delta r\tilde{l}(1+2\tilde{l})}{\sqrt{1+\tilde{l}}}-\frac{\epsilon\partial_t^2\delta r}{\sqrt{1+\tilde{l}}},\nonumber\\
&&K_{ta}=-\frac{\epsilon\partial_a\partial_t\delta r}{\sqrt{1+\tilde{l}}},\nonumber\\
&&K_{ab}=\sqrt{1+\tilde{l}}\Omega_{ab}+\frac{\epsilon\delta r(1+2\tilde{l})}{\sqrt{1+\tilde{l}}}\Omega_{ab}-\frac{\epsilon\nabla^{(s)}_a\nabla^{(s)}_b\delta r}{\sqrt{1+\tilde{l}}}.
\end{eqnarray}
The trace of the extrinsic curvature $\mathcal{K}$ can as well be evaluated in the bulk at the membrane surface and its expression is given by
\begin{eqnarray}
\mathcal{K}&&=\left(g^{tt}\nabla_t n_t+r^{rr}\nabla_rn_r+g^{ab}\nabla_a n_b\right)|_{r=1+\epsilon\delta r},\nonumber\\
 &&=\frac{\epsilon \partial^2_t \delta r}{(1+\tilde{l}^{3/2})}+\frac{\tilde{l}}{\sqrt{1+\tilde{l}}}\bigg(1+\epsilon \frac{\delta r}{1+\tilde{l}}\bigg)+(D-2)\sqrt{1+\tilde{l}}\left(1-\frac{\epsilon\delta r}{1+\tilde{l}}\right)-\frac{\epsilon{\nabla^{(s)}}^2\delta r}{\sqrt{1+\tilde{l}}},\nonumber
\end{eqnarray}
where, $\nabla^{(s)}$ denotes gradient w.r.t the unit sphere coordinates.

The induced metric on the membrane world-volume is given by
\begin{equation}\label{aindbh}
ds_{ind}^2=-dt^2(1+\tilde{l}+2\epsilon \tilde{l}\delta r)+(1+2\epsilon \delta r) d\Omega^2_{D-2}.
\end{equation}
Since, the background configuration is static the background velocity field points in the $dt$ direction. The unit normalised background velocity field is given by 
$$u=-\sqrt{1+\tilde{l}} dt,$$ and the fluctuations about it can be parametrised as
\begin{equation}
u=-\sqrt{1+\tilde{l}}dt+\epsilon \delta u_t(t,\theta^a)dt+\epsilon \delta u_a(t,\theta^a)d{\theta}^a.
\end{equation}
Imposing the condition that the above velocity field is unit normalised w.r.t the induced metric up to linear order in the amplitude fluctuation parameter $\epsilon$ gives 
 $$ \delta u_t(t,\theta^a)=-\frac{\tilde{l}\delta r}{\sqrt{1+\tilde{l}}}.$$

The scalar membrane equation $\nabla\cdot u=0$ then evaluates to 
\begin{eqnarray}
(D-2)\frac{\partial_t\delta r}{\sqrt{1+\tilde{l}}}+\nabla_a^{(s)}\delta u^a=0.
\end{eqnarray}
We can write the vector membrane equation as $$V_\mu=\tilde{V}_\nu\mathcal{P}^\nu_\mu,$$ where 
$$\tilde{V}_\mu= \bigg(\frac{\nabla^2 u_\mu}{\mathcal{K}}+u^\nu K_{\mu\nu}-\frac{\nabla_\mu  \mathcal{K}}{\mathcal{K}}\bigg(1-\tilde{\kappa}a_3 \frac{\mathcal{K}^2}{(D-3)^2}\bigg)-(u.\nabla) u_\mu\bigg(1+\tilde{\kappa}a_3\frac{\mathcal{K}^2}{(D-3)^2}\bigg)\bigg).$$
We evaluate the non-zero components of the projector orthogonal to the $u$ vector in the world-volume of the membrane given by $$\mathcal{P}^\mu_\nu=\delta^\mu_\nu+u^\mu u_\nu,$$ to be
\begin{eqnarray}
&&\mathcal{P}_t^t=0,\quad \quad \mathcal{P}^a_t=-\epsilon \sqrt{1+\tilde{l}}\delta u^a,\nonumber\\
&&\mathcal{P}^t_a=\frac{\epsilon\delta u_a}{\sqrt{1+\tilde{l}}}, \quad\quad \mathcal{P}_a^b=\delta_a^b.
\end{eqnarray} 
From above it is clear that the membrane equation can have non-trivial terms along the $t$ direction at order $\mathcal{O}(\epsilon)$ only if there are terms in $\tilde{V}_a$ that are order $\mathcal{O}(1)$. We find that all terms in $\tilde{V}_a$ are order $\mathcal{O}(\epsilon)$ and hence the vector membrane equation has non-trivial components only along the $\theta^a$ directions. 

The terms relevant for the computation of the vector membrane equation are given below. We only keep terms which contribute to the vector equation at leading order in large $D$,
\begin{eqnarray}
u\cdot\nabla u_t&=&0, \nonumber\\
u\cdot\nabla u_a&=&\frac{\epsilon}{\sqrt{1+\tilde{l}}}\left(\partial_t\delta u_a+\frac{\tilde{l}\partial_a\delta r}{\sqrt{1+\tilde{l}}}\right),\nonumber\\
(u.K)_t &=&-\tilde{l}+\mathcal{O}(\epsilon),\nonumber\\
(u.K)_a &=& -\frac{\epsilon\partial_t\partial_a \delta r}{1+\tilde{l}}+\epsilon\delta u_a\sqrt{1+\tilde{l}},\nonumber\\
\nabla_t \mathcal{K} &=& \mathcal{O}(\epsilon),\nonumber\\
\nabla_a \mathcal{K} &=& \frac{\epsilon}{\sqrt{1+\tilde{l}}}\left(\frac{\partial^2_t\partial_a \delta r}{1+\tilde{l}}+\frac{\tilde{l}\partial_a \delta r}{1+\tilde{l}}-\partial_a {\nabla^{(s)}}^2\delta r-(D-2)\partial_a \delta r\right),\nonumber\\
\frac{\nabla_a\mathcal{K}}{\mathcal{K}}&=&-\frac{\epsilon \nabla_a}{1+\tilde{l}}\left(1+\frac{{\nabla^{(s)}}^2}{D}\right)\delta r\nonumber\\
\nabla^2 u_t &=& 0,\nonumber\\
\nabla^2 u_a &=& -\frac{\epsilon}{(1+\tilde{l})^{3/2}}\left(\sqrt{1+\tilde{l}}\partial^2_t\delta u_a+\tilde{l}\partial_t \partial_a \delta r\right)+ \epsilon {\nabla^{(s)}}^2 \delta u_a+\frac{\epsilon}{\sqrt{1+\tilde{l}}}\nabla_a \delta r\nonumber\\
\frac{\nabla^2 u_a}{\mathcal{K}} &=& \epsilon \frac{{\nabla^{(s)}}^2 \delta u_a}{(D-2)\sqrt{1+\tilde{l}}}.
\end{eqnarray}
Collecting the expressions written above, the
component of the linearised vector membrane equation along the angular direction is given by
\begin{eqnarray}\label{vmembb}
V_a
 &=& \frac{\epsilon}{\sqrt{1+\tilde{l}}}\left(\frac{{\nabla^{(s)}}^2 \delta u_a}{D+2}-(1+\tilde{\kappa}a_3(1+\tilde{l}))\partial_t \delta u_a+(1+\tilde{l})\delta u_a\right)-\frac{\epsilon \tilde{l}\delta u_a}{\sqrt{1+\tilde{l}}} \nonumber\\&&+\frac{\epsilon}{1+\tilde{l}}\Bigg((1-\tilde{\kappa}a_3(1+\tilde{l}))\left(\frac{\nabla^{(s)}_a{\nabla^{(s)}}^2 \delta r}{D+2}+\nabla^{(s)} _a \delta r\right)-(1+\tilde{\kappa}a_3(1+\tilde{l}))\tilde{l}\nabla^{(s)}_a \delta r \nonumber\\ &&-\partial_t\nabla^{(s)}_a \delta r)\Bigg).
\end{eqnarray}
The vector equation above can be decomposed into a part which is the gradient of a scalar function and pure divergenceless vector part as given below
\begin{eqnarray}
V_a=E^{(V)}_a+\nabla^{(s)}_a E^{(S)}=0
\end{eqnarray}
Once can take a divergence of the above equation to get
$$\nabla_{(s)}^2E^{(S)}=0$$
which in the configuration that we are working with evaluates to
\begin{eqnarray}\label{diveq}
&&\frac{\epsilon}{\sqrt{1+\tilde{l}}}\left(\frac{{\nabla^{(s)}}^a\nabla^2 \delta u_a}{D+2}-(1+\tilde{\kappa}a_3(1+\tilde{l}))\partial_t {\nabla^{(s)}}^a\delta u_a
+(1+\tilde{l}){\nabla^{(s)}}^a\delta u_a-\tilde{l}{\nabla^{(s)}}^a\delta u_a\right)\nonumber\\
&&+\frac{\epsilon}{1+\tilde{l}}\Bigg((1-\tilde{\kappa}a_3(1+\tilde{l}))\left(\frac{{\nabla^{(s)}}^2{\nabla^{(s)}}^2 \delta r}{D}+{\nabla^{(s)}}^2 \delta r\right)
-(1+\tilde{\kappa}a_3(1+\tilde{l}))\tilde{l}{\nabla^{(s)}}^2 \delta r \nonumber \\ &&-\partial_t {\nabla^{(s)}}^2 \delta r \Bigg)=0.
\end{eqnarray}
We can decompose the angular component of the velocity vector field also into a gradient of a scalar part and a divergenceless vector part as $$\delta u_a=\delta v_a+\nabla^{(s)}_a \Phi, \quad \text{with}\quad  {\nabla^{(s)}}^a\delta v_a=0,$$
This together with the scalar membrane equation turns the divergence of the vector equation into an equation for $\delta r$ given by
\begin{eqnarray}\label{bhscal}
\nonumber &&-{\nabla^{(s)}}^2 \delta r+(1+\tilde{\kappa}a_3(1+\tilde{l}))(D-2) \partial_t \delta r-2(D-2)\partial_t \delta r
 \\&&+(1-\tilde{\kappa}a_3(1+\tilde{l}))\bigg(\frac{{\nabla^{(s)}}^2{\nabla^{(s)}}^2\delta r}{D}+{\nabla^{(s)}}^2 \delta r \bigg)
-(1+\tilde{\kappa}a_3(1+\tilde{l}))\tilde{l}{\nabla^{(s)}}^2\delta r\nonumber \\&&-\partial_t {\nabla^{(s)}}^2 \delta r=0.
\end{eqnarray}
The background has an $SO(D-2)\times R^{(1)}$ isometry and all the functions are characterised by their decomposition into spherical harmonics and the frequency in each spherical harmonic sector e.g.
 $$ \delta r=\sum_{j,m} c_{jm}Y_{jm}e^{-i w^s_j t},$$  where the spherical harmonics obey $$ {\nabla^{(s)}}^2 Y_{jm}=-j(D+j-3)Y_{jm} .$$

Substituting the above decomposition of $\delta r$ into the equation for $\delta r$  gives the spectrum of the scalar fluctuations to be
\begin{eqnarray}
w^s_j&=&\pm \left(\frac{1}{\sqrt{j(1+l)-1}}\Big(\big(1-a_3\tilde{\kappa}(1+l)\big)(j-1)+j l \Big)+\frac{a_1\tilde{\kappa}j l^2}{2\sqrt{j(1+l)-1}}\right)\nonumber \\
&&+i (1-a_3 \tilde{\kappa}(1+l))(1-j).
\end{eqnarray}
Once, the above equation is satisfied we have $\nabla^2_{(s)}E^{(s)}=0$. On a sphere this is true iff $E^{(s)}=0$ and hence we need to solve the equation $E^{(V)}=0$ i.e. the divergenceless part of the vector membrane equation along the spherical coordinates only. This equation is obtained by taking the part of the equation containing $\delta u_a$ and replacing $\delta u_a$ with $\delta v_a$. Following this procedure we get the equation 
\begin{equation}
 \frac{\epsilon}{\sqrt{1+\tilde{l}}}\left(\frac{{\nabla^{(s)}}^2 \delta v_a}{D+2}-(1+\tilde{\kappa}a_3(1+\tilde{l}))\partial_t \delta v_a+(1+\tilde{l})\delta v_a\right)-\frac{\epsilon \tilde{l}\delta v_a}{\sqrt{1+\tilde{l}}}=0.
\end{equation}

Substituting the decomposition of $\delta v_a$ into vector spherical harmonics sectors given by $\delta v_a= \sum_{j,m} d_{j,m}Y_a^{j,m} e^{-iw_v^j t}$ into the above equation gives the spectrum of the 
vector quasi-normal mode frequency as
\begin{equation}
w_v=i (1-a_3 \tilde{\kappa}(1+l))(1-j).
\end{equation}
These results match with the scalar and vector quasi-normal mode frequencies of $AdS$ black hole found in \cite{Emparan:2015rva} and \cite{Bhattacharyya:2017hpj} in the limit $\tilde{\kappa}\to 0$. For Gauss-Bonnet combination i.e. $a_1=1, a_2=-4$ and $a_3=1$, the quasi-normal mode frequencies agree with \cite{Chen:2015fuf,Chen:2017hwm} in the appropriate limit.

It is easy to see that the mode $j=0$ is a zero mode for the scalar mode. This can be seen both from the expression of the frequencies of the spectrum given above and also from the fact that the decoupled equation for $\delta r$ is trivial for $j=0$ and the scalar membrane equation for $j=0$ gives $w=0$. This mode corresponds to a uniform scaling of the size of the black hole and is obviously a solution of the gravity equations. There are no zero modes corresponding to $j=1$ scalar mode. The presence of this zero mode would have implied that translations and boost of the centre of mass of the black hole are solutions of the gravity equation. This is not the case here as $AdS$ spacetime has a non-trivial warping. 

Similarly, the vector sector has a zero mode at $j=1$ and corresponds to uniform rotation of the black hole about an axis, which also is obviously a solution of the gravity equations. 
\subsection{Quasi-normal mode of a black brane }
In this subsection, we compute the frequency of linearised fluctuations about a membrane dual to a $AdS$ black brane in the four-derivative gravity theory under consideration. The asymptotic spacetime of the black brane in which the membrane propagates is given by
\begin{equation}\label{bbbcg}
ds^2 = -\tilde{l} r^2dt^2+\frac{dr^2}{\tilde{l}r^2}+l r^2 dx^2_{D-2},
\end{equation}
where, $\tilde{l}=l(1+\tilde{\kappa}\tilde{\mathcal{A}}l)$. 

We consider a small amplitude fluctuation around the static membrane solutions dual to black brane $r=1$ parametrised by
\begin{equation}
r=1+\epsilon \delta r(t,x^a).
\end{equation}
The unit normal to this membrane surface is given by
\begin{eqnarray}
n &=& \frac{dr-\epsilon \partial_t \delta r dt-\epsilon \partial_a\delta rdx^a}{\sqrt{\tilde{l}r^2}}.
\end{eqnarray}

Non-trivial components of the pullback of the spacetime extrinsic curvature on the membrane $r=1+\epsilon \delta r$ are given by
\begin{eqnarray}
K_{tt} &=& -\epsilon \frac{\partial^2_t \delta r}{\sqrt{\tilde{l}}}-\tilde{l}^{3/2}(1+2\epsilon \delta r), \quad K_{ta} = -\epsilon \frac{\partial_t\partial_a \delta r}{\sqrt{\tilde{l}}}, \nonumber \\
K_{ab} &=& -\epsilon \frac{\partial_a\partial_b \delta r}{\sqrt{\tilde{l}}}+l\sqrt{\tilde{l}}(1+2\epsilon \delta r)\delta_{ab}.
\end{eqnarray}
The trace of the extrinsic curvature on the membrane surface takes the form $$\mathcal{K}=\epsilon\frac{\partial^2_t \delta r}{\tilde{l}^{3/2}}-\epsilon \frac{\partial^2 \delta r}{l\tilde{l}}+(D-1)\sqrt{\tilde{l}},\quad \text{where} \quad \partial^2=\partial^a\partial_a.$$
The induced metric on the membrane is
\begin{equation}\label{lind}
ds^2 =-\tilde{l}(1+2\epsilon \delta r) dt^2+l(1+2\epsilon \delta r)dx^2_{D-2}.
\end{equation}
We parametrise small fluctuation around static velocity configuration $u=-\sqrt{\tilde{l}r^2}dt$  by
\begin{equation}\label{linvel}
u=-\sqrt{\tilde{l}}dt+\epsilon \delta u_t(t,a) dt+\epsilon \delta u_a(t,a) dx^a.
\end{equation}
Imposing the condition that $u.u=-1$ w.r.t the induced metric \eqref{lind}, we find $\delta u_t(t,a)= -\sqrt{\tilde{l}} \delta r(t,a)$.

Non-zero components of the projector orthogonal to $u$ are given by
\begin{eqnarray}
P^t_t &=& 0,\quad P^t_a=\epsilon \frac{\delta u_a}{\sqrt{\tilde{l}}}, \quad P^a_t= -\epsilon \frac{\sqrt{\tilde{l}}}{l}\delta u^a \quad \text{and} \quad p^a_b=\delta^a_b.
\end{eqnarray}
Explicit expressions for the terms appearing in the vector membrane equation are given by
\begin{eqnarray}
\nabla^2u_t &=& 0, \nonumber \\
 \nabla^2 u_a&=&\epsilon\left(-\frac{1}{\tilde{l}}\partial^2_t \delta u_a+\frac{1}{l}\partial^b\partial_b \delta u_a\right),\nonumber \\
(u.K)_t &=& -\epsilon \left(\frac{1}{\tilde{l}}\partial^2_t \delta r+\tilde{l}\delta r\right)-\tilde{l},\nonumber\\
(u.K)_a &=& -\epsilon\left(\frac{\partial_t \partial_a \delta r}{\tilde{l}}-\sqrt{\tilde{l}}\delta u_a \right),\nonumber\\
\nabla_t \mathcal{K} &=& \epsilon \left( \frac{\partial^3_t \delta r}{\tilde{l}^{3/2}}-\frac{\partial_t \partial^2 \delta r }{l\sqrt{\tilde{l}}}\right),\nonumber\\
\nabla_a \mathcal{K} &=& \epsilon \left( \frac{\partial_a\partial^2_t \delta r}{\tilde{l}^{3/2}}-\frac{\partial_a \partial^2 \delta r }{l\sqrt{\tilde{l}}}\right),\nonumber\\
(u.\nabla)u_t &=& 0,\nonumber \\
(u.\nabla)u_a &=& \epsilon\left(\frac{1}{\sqrt{\tilde{l}}}\partial_t \delta u_a+\partial_a \delta r \right).
\end{eqnarray}
Here also the vector membrane equation has non-zero component at $\mathcal{O}(\epsilon)$ only along the $x^a$ directions. Using expressions evaluated above, this component of the equation is given by 
\begin{eqnarray}\label{bbmem}
&&\frac{1}{\tilde{l}}\partial_a\partial_t \delta r+(1+\tilde{\kappa}a_3 \tilde{l})\partial_a \delta r+\frac{1-\tilde{\kappa}a_3 \tilde{l}}{D}\left(\frac{\partial_a \partial^2_t \delta r}{\tilde{l}^2}-\frac{\partial_a \partial^2 \delta r}{l\tilde{l}}\right)
+\frac{1}{D}\left(\frac{\partial^2_t \delta u_a}{\tilde{l}^{3/2}}-\frac{\partial^2 \delta u_a}{l\tilde{l}^{1/2}}\right)\nonumber\\&&+(1+\tilde{\kappa}a_3 \tilde{l}) \frac{\partial_t \delta u_a}{\tilde{l}^{1/2}}=0.
\end{eqnarray}
The scalar membrane equation $\nabla.u=0$ evaluates to linear order in $\epsilon$ to
\begin{equation}\label{ucond}
\frac{\partial^a\delta u_a }{l}+\frac{(D-2)}{\sqrt{\tilde{l}}}\partial_t \delta r=0.
\end{equation}
We decompose the shape and velocity fluctuations according to the symmetry of the background configuration in terms of Fourier transform in momentum basis along the $x^a$ directions and frequency basis in the  time direction 
\begin{eqnarray}
\delta r&=&\delta r^0_{k,w}e^{-i w t}e ^{i k_a x^a},\nonumber\\
\delta u_a&=&\delta u_{a,k,w}^0e^{-i w t}e ^{i k_a x^a}.
\end{eqnarray}
The above equation can be solved in two different scaling limits as we now explain
\subsection{$\mathcal{O}(1)$ derivatives}
In this scaling we assume that all the spatial and temporal derivatives are $\mathcal{O}(1)$. Hence, we drop all subleading terms from the vector membrane equation to get
\begin{eqnarray}\label{ref1}
&&\frac{1}{\tilde{l}}\partial_a\partial_t \delta r+(1+\tilde{ \kappa}a_3 \tilde{l})\partial_a\delta r+(1+\tilde{\kappa}a_3 \tilde{l}) \frac{\partial_t \delta u_a}{\tilde{l}^{1/2}}=0.
\end{eqnarray}
Taking the divergence of the above equation in the $x^a$ directions and substituting the scalar membrane equation into it we get
\begin{eqnarray}
&&\frac{1}{\tilde{l}}\partial^2\partial_t \delta r+(1+\tilde{ \kappa}a_3 \tilde{l})\partial^2\delta r-D\frac{l}{\tilde{l}}(1+\tilde{\kappa}a_3 \tilde{l}) \partial_t^2\delta r=0.
\end{eqnarray}
The solution to this equation is 
\begin{eqnarray}
&&\delta r=\delta r_{w_s^1}e^{-i w_s^1+i k_a x^a}+\delta r_{w_s^2}e^{-i w_s^2+i k_a x^a},\nonumber\\
&&w_s^1=\left(1+\frac{\tilde{ \kappa}}{2} a_1 l\right)\frac{k}{\sqrt{D}},\nonumber\\
&&w_s^2=-\left(1+\frac{\tilde{ \kappa}}{2} a_1 l\right)\frac{k}{\sqrt{D}}.
\end{eqnarray}
Substituting this into \eqref{ref1} along with the most general form of the solution of the velocity fluctuations given by $$\delta u_a =V_a^1e^{-i w_s^1+i k_a x^a}+V_a^2e^{-i w_s^2+i k_a x^a}+v_a e^{-i w_v+i k_a x^a},$$where, $v_a k^a=0$ gives
\begin{eqnarray}
&& V_a^1=\left(\frac{-i}{w_s^1}\frac{(1-\tilde{ \kappa} a_3 \tilde{l})}{\sqrt{\tilde{l}}}k_a w_s^1 \delta +\frac{\sqrt{\tilde{l}}}{w_s^1}k_a\right)\delta r_1,\nonumber\\
&&V_a^2=\left(\frac{-i}{w_s^2}\frac{(1-\tilde{ \kappa} a_3 \tilde{l})}{\sqrt{\tilde{l}}}k_a w_s^2 \delta +\frac{\sqrt{\tilde{l}}}{w_s^2}k_a\right)\delta r_1,\nonumber\\
&&w_v=\mathcal{O}(1/D).
\end{eqnarray}
The scalar quasi-normal modes defined above look like normal-modes. This is odd since the corresponding problem in the gravity picture has a leaking boundary condition at the horizon of the black brane. The $\sqrt{D}$ dependence of the solution together with the observations made in \cite{Dandekar:2016jrp} and \cite{Bhattacharyya:2017hpj} points to the fact that working in scaling limit where $k_a\sim\mathcal{O}(\sqrt{D})$ gives interesting results for the quasi-normal modes. 
\subsection{$\mathcal{O}(\sqrt{D})$ spatial derivatives}
Taking the divergence of the vector membrane equation \eqref{veccon} and substituting the  expression for $\partial^a\delta u_a$ from the scalar membrane equations gives 
\begin{eqnarray}\label{bbdiv}
&&\epsilon\Bigg(\frac{1}{\tilde{l}}\partial^2\partial_t \delta r+(1+\tilde{\kappa}a_3 \tilde{l})\partial^2 \delta r+\frac{1-\tilde{\kappa}a_3 \tilde{l}}{D}\left(\frac{\partial^2 \partial^2_t \delta r}{\tilde{l}^2}-\frac{\partial^2 \partial^2 \delta r}{l\tilde{l}}\right)
-\bigg(\frac{l \partial^3_t\delta r }{\tilde{l}^2}-\frac{\partial^2\partial_t\delta r}{\tilde{l}}\bigg)\nonumber\\&&-D(1+\tilde{\kappa}a_3 \tilde{l}) \frac{l\partial_t^2 \delta r}{\tilde{l}}\Bigg)=0.
\end{eqnarray}
We will work in a limit where the momenta $k_a\sim\mathcal{O}(\sqrt{D})$ and hence we write $k_a= q_a\sqrt{D}$ where, $q_a$ is $\mathcal{O}(1)$. The derivatives in the time direction are still $\mathcal{O}(1)$. In this scaling limit the terms in the above equation that contribute at leading and subleading order in large $D$ limit are
\begin{eqnarray}
\frac{1}{\tilde{l}}\partial^2\partial_t \delta r+(1+\tilde{\kappa}a_3 \tilde{l})\partial^2 \delta r-\frac{1-\tilde{\kappa}a_3 \tilde{l}}{D}\frac{\partial^2 \partial^2 \delta r}{l\tilde{l}}
+\frac{\partial^2\partial_t\delta r}{\tilde{l}}-D(1+\tilde{\kappa}a_3 \tilde{l}) \frac{l\partial_t^2 \delta r}{\tilde{l}}=0.\nonumber\\
\end{eqnarray}
We keep the first subleading parts to get a non-zero answer for the frequency. Substituting $\delta r= \delta r_0 e^{-i w_s t}e^{i k_a x^a}$  in \eqref{bbdiv}, we arrive at the spectrum of the scalar fluctuations
\begin{eqnarray}
&&\delta r= \delta r_0^1 e^{-i w^1_s t}e^{i k_a x^a}+ \delta r_0^2 e^{-i w^2_s t}e^{i k_a x^a},\nonumber\\
&&w_s^1 = \left(1+\tilde{\kappa}a_1 \frac{l}{2}\right)q-i \left(1-\tilde{\kappa}a_3 l\right)\frac{q^2}{l}, \nonumber\\
&&w_s^2=- \left(1+\tilde{\kappa}a_1 \frac{l}{2}\right)q-i \left(1-\tilde{\kappa}a_3 l\right)\frac{q^2}{l},\quad \text{where} \quad  q=\sqrt{q^aq_a}=\frac{\sqrt{k^ak_a}}{\sqrt{D}}.\nonumber\\
\end{eqnarray}

We again consider the most general form of $\delta u_a$ 
\begin{eqnarray}\label{bbvec}
&&\delta u_a = V^1_a \delta r_0^1 e^{-i w^1_s t}e^{i k_a x^a}+V^2_a \delta r_0^2 e^{-i w^2_s t}e^{i k_a x^a}+v_a e^{-i w_v t}e^{i k_a x^a},\nonumber\\
&&\text{where,}\quad k^av_a=0,
\end{eqnarray}
and substitute it in the vector membrane equation to get
\begin{eqnarray}
V^1_a &=& \left(\frac{\sqrt{l D}}{k}-\frac{i}{\sqrt{l}}\left(1-\tilde{\kappa}a_3l-\tilde{\kappa}a_1 \frac{l}{2}\right)\right)k_a,\nonumber\\
V^2_a &=& \left(-\frac{\sqrt{l D}}{k}-\frac{i}{\sqrt{l}}\left(1-\tilde{\kappa}a_3l-\tilde{\kappa}a_1 \frac{l}{2}\right)\right)k_a,
\end{eqnarray}
and the frequency of the vector fluctuations to be
\begin{eqnarray}
w_v &=& -i\frac{q^2}{l }\left(1-\tilde{\kappa}a_3l\right).
\end{eqnarray}
In the absence of higher derivative correction, these results agree with \cite{Bhattacharyya:2017hpj}.

\section{Outlook and future directions}\label{con}
The main result of this paper is the computation of the membrane equations dual to the dynamics of black holes in the most general four-derivative theory of gravity to leading order in a $1/D$ expansion. Like in \cite{Saha:2018elg} we have worked to linear order in the parameter specifying the relative strength of the four-derivative terms in the action w.r.t the two-derivative Einstein-Hilbert part of the action. The $AdS$ radius of the asymptotic spacetime of the solutions is modified in presence of the higher derivative terms as $L_{AdS}^2=\frac{1}{l(1+\tilde{\kappa}a_1 l )}.$
The effective equation of the membrane when expressed in terms of quantities expressed in this $AdS$ spacetime is the covariantisation of the corresponding membrane equation dual to black holes dynamics in asymptotic flat spacetime for the four-derivative theory of gravity. Together with the observation made for Einstein-Hilbert gravity in \cite{Bhattacharyya:2017hpj} this can be seen as a general property of membranes dual to large $D$  dynamics of black holes in absence of charge: To leading order in large $D$ membranes dual to black holes in asymptotic $AdS$ spacetime have equations which are covariant versions of the corresponding equations in flat spacetime. We do not see any obvious reason for this to be true and we think that this needs to be understood better. 

Another interesting thing that we observed is the map between membranes dual to four-derivative theory of gravity up to linear order in $\tilde{ \kappa}$ and those dual to charged black holes \cite{Bhattacharyya:2015fdk, Kundu:2018dvx}, continue to persist even in the presence of the cosmological constant. Under the map $$Q\rightarrow \sqrt{\tilde{ \kappa}a_3}\frac{\mathcal{K}}{D},$$
the charged membrane equations map exactly to the membrane equations derived here. In fact the non-trivial appearance of the curvature of the background spacetime in the charge case appears non-trivially in the divergence of the vector membrane equation here via the Gauss-Codacci equations. 

Since, there is no known independent theory of membranes which can be dual to black hole dynamics in general, the above two observations can be thought of as data points to get better intuition about  the ``transport coefficients'' of the membranes. It will be interesting to extend the analysis to find the dual membranes in more general backgrounds to get a better understanding of the structure of the membrane equations and stress tensor. A first step in this direction in the mass and momentum approach has been taken in \cite{Andrade:2018zeb}. The authors of this paper derived the effective equations in $AdS$ in presence of arbitrary boundary deformation with particular large $D$ scaling. Due to the choice of this scaling they observe the presence of the effect of curvature of the asymptotic metric even at leading order.  It will be interesting to study similar phenomena in the membrane picture  as this may help us in formulating the effect of background curvature on the membrane. 

It will also be interesting to extend the analysis to subleading order both in $\tilde{ \kappa}$ and in $1/D$. The analysis at subleading order in $\tilde{ \kappa}$ will help us to understand if the phenomenon of covariantisation of the membrane equation still persists or not. The analysis at subleading order in $1/D$ will be important from the point of view of collecting more data about the second order membrane equations which played crucial role in the derivation of the entropy current for two-derivative theory of gravity.  






 \section*{Acknowledgement}
 We would like to thank N. Banerjee for collaboration during the initial stages of this work and many insightful discussions at various stages. A.K would like to acknowledge hospitality of IISER Bhopal. T.M. would like to thank D. Mukherjee, B. Chakrabarty, P. Biswas for many helpful discussions. A.S. would like to thank Y. Dandekar for a discussion related to the computation of quasi-normal modes of black branes.  T.M. was partially supported by SERB, DST `Ramanujan Fellowship' of Nabamita Banerjee at IISER Pune. A.S. is supported by the Ambizione grant no. $PZ00P2\_174225/1$ of the Swiss National Science Foundation (SNSF) and partially supported by the NCCR grant no. $51NF40-141869$ ``The Mathematics of Physics'' (SwissMAP)

\appendix

 \section{Black brane solutions}\label{leadans}
 
 
 The static black brane solution to \eqref{eom} (perturbatively in $\kappa$) is given by \cite{Kats:2007mq}
 \begin{equation}\label{imet}
 ds^2= -f(r)d\tilde{t}^2+\frac{dr^2}{f(r)}+l r^2 d x_{D-2}^2,
 \end{equation}
 where \begin{equation}
 f(r)=l r^2\Bigg(1-\Big(\frac{r_h}{r}\Big)^{D-1}(1+\tilde{\kappa} l(\tilde{\mathcal{A}}+a_3))+\tilde{\kappa} l \tilde{\mathcal{A}}+\tilde{\kappa}la_3\Big(\frac{r_h}{r}\Big)^{2(D-1)}\Bigg),
 \end{equation}
 and 
 \begin{eqnarray}
 \tilde{\mathcal{A}}&=&\frac{1}{(D-3)(D-2)}\Big((D-1)(D a_1+a_2)+2a_3\Big),\\
 \tilde{\kappa}&=&(D-4)(D-3)\kappa.
 \end{eqnarray}
 $r=r_h$ is the position of the horizon. The effect of the large $D$ limit can be easily understood if we write the solution in Kerr-Schild coordinate. We introduce a new coordinate $V$ such that : $d\tilde{t}= \frac{dV}{\sqrt{l r^2(1+\tilde{\kappa}\tilde{\mathcal{A}}l)}}-\frac{dr}{f(r)}$ and the metric \eqref{imet} takes the following form,
 \begin{equation}\label{smet}
 ds^2= -\frac{f(r)dV^2}{l r^2(1+\tilde{\kappa}\tilde{\mathcal{A}}l)}+\frac{2dV dr}{\sqrt{l r^2(1+\tilde{\kappa}\tilde{\mathcal{A}}l)}}+lr^2 dx^2_{D-2}.
 \end{equation}
 We again do a coordinate change of the form $dV=\sqrt{l r^2(1+\tilde{\kappa}\tilde{\mathcal{A}}l)}dt+\frac{dr}{\sqrt{l r^2(1+\tilde{\kappa}\tilde{\mathcal{A}}l)}}$, then the solution \eqref{smet} reduces to 
 \begin{eqnarray}\label{kch}
 ds^2&=& -l r^2 (1+\tilde{\kappa}l \tilde{\mathcal{A}})dt^2+\frac{dr^2}{l r^2 (1+\tilde{\kappa}l\tilde{\mathcal{A}})}+l r^2 dx^2_{D-2}\nonumber\\
 &&+\frac{g(r)}{(1+\tilde{\kappa}\tilde{\mathcal{A}}l)}\left(\sqrt{l r^2(1+\tilde{\kappa}\tilde{\mathcal{A}}l)}dt+\frac{dr}{\sqrt{l r^2(1+\tilde{\kappa}\tilde{\mathcal{A}}l)}}\right)^2\nonumber\\
 &=& ds^2_{asym}+\frac{g(r)}{(1+\tilde{\kappa}\tilde{\mathcal{A}}l)}\left(\sqrt{l r^2(1+\tilde{\kappa}\tilde{\mathcal{A}}l)}dt+\frac{dr}{\sqrt{l r^2(1+\tilde{\kappa}\tilde{\mathcal{A}}l)}}\right)^2.
 \end{eqnarray}
 where $$g(r)=\left(\frac{r_h}{r}\right)^{D-1}\left(1+\tilde{\kappa}l(\tilde{\mathcal{A}}+a_3)\right)-\tilde{\kappa}la_3\left(\frac{r_h}{r}\right)^{2(D-1)},$$ and
 \begin{equation}\label{bbcg}
 ds_{asym}^2= -l r^2 (1+\tilde{\kappa}l \tilde{\mathcal{A}})dt^2+\frac{dr^2}{l r^2 (1+\tilde{\kappa}l\tilde{\mathcal{A}})}+l r^2 dx^2_{D-2}.
 \end{equation}
 In the $D\rightarrow \infty$, at $r\gg r_h$, the function $g(r)$ vanishes, which implies that in the large $D$ limit, blackening factor is non-trivial only around a small thickness around the horizon. It is also worth to notice that the background geometry\eqref{bbcg}  is modified in the presence of the higher derivative parameter.

 
 The above metric \eqref{kch} can be written in a more covariant form given by
 \begin{eqnarray}\label{kerr}
 ds^2&=&ds_{asym}^2+\Bigg(\psi^{-(D-1)}\Big(1+\tilde{\kappa}a_3 \frac{\mathcal{K}^2}{(D-2)^2}\Big)-\tilde{\kappa}a_3 \frac{\mathcal{K}^2}{(D-2)^2}\psi^{-2(D-1)}\Bigg)(O_M dx^M)^2,\nonumber\\
 \text{where,}\nonumber\\
 O_Mdx^M&=&n-u, \quad \text{and}\quad n= \frac{dr}{\sqrt{l r^2(1+\tilde{\kappa}a_1 l)}}, u= -\sqrt{l r^2(1+\tilde{\kappa}a_1 l)}dt,
 \end{eqnarray}
 where $\tilde{\kappa}=(D-4)(D-3)\kappa$ and is $\mathcal{O}(1)$ quantity and $\mathcal{K}$ is the trace of the extrinsic curvature of the hypersurface $r=r_h$ embedded in the asymptotic $AdS$ spacetime and is given by $$\mathcal{K}=(D-2) \sqrt{l(1+\tilde{\kappa}\tilde{\mathcal{A}}l)}.$$
 
 Also, $O_M= n_M-u_M$, where $n_Mdx^M=\frac{dr}{\sqrt{l r^2(1+\tilde{\kappa}a_1 l)}}$ is the unit normal to the surface $r=r_h$ (normalised w.r.t the asymptotic $AdS$ spacetime) and $u_Mdx^M=-\sqrt{l r^2(1+\tilde{\kappa}a_1 l)}dt$  such that
 \begin{equation}
 n.n=1 \quad n.u=0 \quad \text{and} \quad \ u.u=-1.
 \end{equation}
 All dot products are taken w.r.t the asymptotic $AdS$ spacetime given by
 \begin{eqnarray}\label{bcgmet}
 ds^2&=&-l r^2 (1+\tilde{\kappa}l \tilde{\mathcal{A}})dt^2+\frac{dr^2}{l r^2 (1+\tilde{\kappa}l\tilde{\mathcal{A}})}+l r^2 dx^2_{D-2} .
 \end{eqnarray}
 
 \section{Details of quasi-normal mode calculation }
 \subsection{$AdS$ black hole}\label{details_qnm_Bhole}
 The non-trivial Christoffel symbols associated with the metric \eqref{abh} are given by
\begin{eqnarray}
&&\Gamma^t_{rt}=\frac{\tilde{l}r}{1+\tilde{l}r^2},\quad \quad\Gamma^r_{rr}=-\frac{\tilde{l}r}{1+\tilde{l}r^2},\nonumber\\
&&\Gamma^r_{tt}=\tilde{l} r (1+\tilde{l}r^2),\quad \quad\Gamma^b_{rd}=\frac{1}{r}\delta^b_d,\nonumber\\
&&\Gamma^r_{ab}=-r\Omega_{ab}(1+\tilde{l}r^2),\quad \quad\Gamma^a_{bc}={{\Gamma}^a}^s_{bc},
\end{eqnarray}
where ${{\Gamma}^a}^s_{bc}$ is the Christoffel symbol of unit $D-2$ sphere.
The covariant derivatives of the unit normal vector with respect to the background metric  \eqref{abh} are given by
\begin{eqnarray}
&&\nabla_t n_t=-\tilde{l} r\sqrt{1+\tilde{l}r^2}-\frac{\epsilon\partial_t^2\delta r}{\sqrt{1+\tilde{l}r^2}},\quad \quad\nabla_tn_r=\frac{\epsilon \tilde{l} r\partial_t\delta r}{(1+\tilde{l}r^2)^{3/2}},\nonumber\\
&&\nabla_t n_a=\frac{-\epsilon \partial_t \partial_a \delta r}{\sqrt{1+\tilde{l}r^2}},\quad \quad\nabla_r n_t=\frac{2\epsilon \tilde{l} r\partial_t\delta r}{(1+\tilde{l}r^2)^{3/2}},\nonumber\\
&&\nabla_r n_a=\epsilon \partial_a \delta r\frac{1+2\tilde{l}r^2}{r(1+\tilde{l} r^2)^{3/2}}, \quad \quad \nabla_r n_r=\frac{-\tilde{l} r}{(1+\tilde{l}r^2)^{3/2}}, \quad \quad\nabla_a n_t=0,\nonumber\\
&&\nabla_a n_r=\frac{\epsilon \partial_a \delta r}{r\sqrt{1+\tilde{l}r^2}},\quad\quad\nabla_a n_b=-\epsilon \frac{{\nabla^{(s)}}_a{ \nabla^{(s)}}_b\delta r}{\sqrt{1+\tilde{l}r^2}}+r \Omega_{ab}\sqrt{1+\tilde{l}r^2}.
\end{eqnarray}
Using the above equations, the components of the spacetime form of extrinsic curvature are given by
\begin{eqnarray}
&&K^s_{tt}=-\tilde{l}r\sqrt{1+\tilde{l}r^2}-\frac{\epsilon \partial_t^2\delta r}{\sqrt{1+\tilde{l}r^2}},\nonumber\\
&&K^s_{tr}=\frac{\epsilon}{2}\frac{\tilde{l} r\partial_t\delta r}{(1+\tilde{l}r^2)^{3/2}},\nonumber\\
&&K^s_{ta}=-\epsilon\frac{\partial_a\partial_t\delta r}{\sqrt{1+\tilde{l}r^2}},\nonumber\\
&&k^s_{rr}=0,\nonumber\\
&&K^s_{ra}=\frac{\epsilon \partial_a\delta r}{2r}\frac{1}{\sqrt{1+\tilde{l}r^2}},\nonumber\\
&&K^s_{ab}=-\epsilon \frac{\nabla^{(s)}_a\nabla^{(s)}_b\delta r}{\sqrt{1+\tilde{l}r^2}}+r\Omega_{ab}\sqrt{1+\tilde{l}r^2}.
\end{eqnarray}
The non-trivial Christoffel symbols of the metric \eqref{aindbh} are given by
\begin{eqnarray}
&&\Gamma^t_{tt}=\frac{\epsilon\tilde{l}\partial_t\delta r}{1+\tilde{l}},\quad\quad\Gamma^t_{ta}=\frac{\epsilon\tilde{l}\partial_a\delta r}{1+\tilde{l}},\nonumber\\
&&\Gamma^t_{ab}=\frac{\epsilon\partial_t\delta r}{1+\tilde{l}}\Omega_{ab},\quad\quad\Gamma^a_{bt}=\epsilon\partial_t\delta r\delta^a_b,\nonumber\\
&&\Gamma^a_{tt}=\epsilon\tilde{l}\Omega^{ab}\partial_b\delta r,\quad\quad\Gamma^a_{bc}=\Gamma^{(s)a}_{bc}+2\epsilon \delta^a_b\partial_c\delta r-\epsilon \Omega^{af}\partial_f\delta r\Omega_{bc}.
\end{eqnarray}
Components of the covariant derivative of the velocity vector w.r.t. the induced metric \eqref{aindbh} are given by
\begin{eqnarray}
&&\nabla_t u_t=0, \quad 
\nabla_a u_b=\epsilon\nabla^{(s)}_a\delta u_b+\epsilon\frac{\partial_t\delta r}{\sqrt{1+\tilde{l}}}\Omega_{ab},\nonumber\\
&&\nabla_a u_t = 0, \quad \nabla_t u_a  = \epsilon \partial_t \delta u_a+\frac{\epsilon \tilde{l}\partial_a \delta r}{\sqrt{1+\tilde{l}}}.
\end{eqnarray}
\subsection{$AdS$ black brane} \label{details_qnm_Bbrane}
The non-zero Christoffel symbols corresponding to the background metric \eqref{bbbcg} are 
\begin{eqnarray}
\Gamma^r_{rr} = -\frac{1}{r},\quad \Gamma^r_{tt}={\tilde{l}}^2r^3, \quad \Gamma^r_{ab}=-l\tilde{l}r^3\delta_{ab}, \quad \Gamma^t_{tr}=\frac{1}{r} \quad \text{and} \quad \Gamma^a_{rb}=\frac{1}{r}\delta^a_b.
\end{eqnarray}
Components of $\nabla_M n_N$ are
\begin{eqnarray}
\nabla_t n_t &=& -\epsilon \frac{\partial^2_t \delta r }{\sqrt{\tilde{l}r^2}}-\tilde{l}^{3/2}r^2, \quad \nabla_t n_r=\epsilon\frac{\partial_t \delta r}{\sqrt{\tilde{l}}r^2} ,\quad \nabla_t n_a= -\epsilon\frac{\partial_t \partial_a\delta r}{\sqrt{\tilde{l}r^2}} ,\nonumber\\
\nabla_r n_t &=&2 \epsilon\frac{\partial_t \delta r}{\sqrt{\tilde{l}}r^2}, \quad \nabla_r n_r=0, \quad \nabla_r n_a =2\epsilon\frac{\partial_a \delta r}{\sqrt{\tilde{l}}r^2},\nonumber \\
\nabla_a n_t &=& -\epsilon\frac{\partial_a \partial_t\delta r}{\sqrt{\tilde{l}r^2}} , \quad \nabla_a n_r=\epsilon\frac{\partial_a \delta r}{\sqrt{\tilde{l}}r^2},\quad\nonumber\\
\nabla_a n_b &=& -\epsilon \frac{\partial_a\partial_b \delta r}{\sqrt{\tilde{l}r^2}}+l\sqrt{\tilde{l}} r^2 \delta_{ab}.
\end{eqnarray}
Components of the spacetime extrinsic curvature are
\begin{eqnarray}
K^s_{tt} &=&  -\epsilon \frac{\partial^2_t \delta r }{\sqrt{\tilde{l}r^2}}-\tilde{l}^{3/2}r^2, \quad  K^s_{tr}=\epsilon\frac{\partial_t \delta r}{\sqrt{\tilde{l}}r^2}, \quad K^s_{ta}= -\epsilon\frac{\partial_t \partial_a\delta r}{\sqrt{\tilde{l}r^2}}, \nonumber\\
K^s_{rr}&=&0, \quad K^s_{ra}=\epsilon\frac{\partial_a \delta r}{\sqrt{\tilde{l}}r^2}\quad\text{and} \quad K^s_{ab}= -\epsilon \frac{\partial_a\partial_b \delta r}{\sqrt{\tilde{l}r^2}}+l\sqrt{\tilde{l}} r^2 \delta_{ab}.
\end{eqnarray}
Christoffel symbols of the induced metric \eqref{lind} are
\begin{eqnarray}
\Gamma^t_{tt} &=& \epsilon \partial_t \delta r, \quad \Gamma^t_{ta}=\epsilon \partial_a \delta r, \quad \Gamma^t_{ab}= \epsilon\frac{l}{\tilde{l}} \partial_t \delta r \delta_{ab}, \quad \Gamma^a_{tt} =\epsilon \frac{\tilde{l}}{l}\partial^a \delta r,\nonumber \\
\Gamma^a_{bt} &=& \epsilon \partial_t \delta r\delta^a_b \quad \text{and} \quad \Gamma^a_{bc} = \epsilon\left(\partial_b \delta r\delta^a_c+\partial_c \delta r\delta^a_b-\partial^a \delta r\delta_{bc}\right).
\end{eqnarray}
Covariant derivatives of various components of the linearized velocity \eqref{linvel}  w.r.t the induced spacetime \eqref{lind} are
\begin{eqnarray}
\nabla_t u_t&=& 0, \quad \nabla_t u_a  = \epsilon \left(\partial_t \delta u_a +\sqrt{\tilde{l}}\partial_a\delta r\right),\quad \nabla_a u_t=0, \nonumber\\
\nabla_a u_b &=& \epsilon\left(\partial_a \delta u_b+\frac{l}{\sqrt{\tilde{l}}}\partial_t \delta r \delta_{ab}\right).
\end{eqnarray}
\section{Map between metric corrections in geometric and effective spacetime}\label{geo_to_eff_map}
This appendix can be thought of as a generalisations of the explicit map worked out in Appendix B of \cite{Bhattacharyya:2015fdk}.
The projector in full spacetime orthogonal to the normal and velocity vector can also be expressed as
\begin{eqnarray}
P_{AB}=p_{AB}+\frac{z_A z_B}{(l r_0^2)^{-1}-n_S^2}+l r_0^2 S_0^2 \Omega_{AB}
\end{eqnarray}
where, $p^A_B$ is the projector orthogonal to $n$, $u$ and the $z$ direction in the effective spacetime and $\Omega_{AB}$ is the unit sphere metric along the ismometry directions and it has non-zero component only along the $\theta_i$ directions specifying the isometry directions. 

We will now find the map between the geometric form of the metric correction of \eqref{geometric_met} and the correction written in the effective spacetime of \eqref{effective_space_metric}. Let us start with the metric correction $H^{(V)}_M$. The metric correction due to this is given by
\begin{eqnarray}
H^{(V)}_M dx^M&=& H^{(V)}_AP^A_M dx^M\nonumber\\
&=& H^{(V)}_A\left(p^A_M+\frac{z^A z_M}{(l r_0^2)^{-1}-n_S^2}+\Omega^A_M\right) dx^M\nonumber\\
&=& H^{(V)}_Ap^A_Mdx^M+H^{(V)}_A z^A\frac{dz}{(l r_0^2)^{-1}-n_S^2}.
\end{eqnarray}
Comparing with the metric correction of \eqref{effective_space_metric} we get
\begin{eqnarray}
H^{(V)}_A p^A_M=V_{Vi}+\tilde{ \kappa}V_{Vi,\tilde{ \kappa}}\quad \text{and}\quad \frac{H^{(V)}_A z^A}{(l r_0^2)^{-1}-n_S^2}=S_{Vz}+\tilde{ \kappa} S_{Vz,\tilde{ \kappa}}.
\end{eqnarray}
The above map is true because the solution preserves an isometry $SO(D-p-3)$ and hence $H^{(V)}_A$ cannot have a component along the isometry direction. 
Since, the solution written in the geometric way also preserves an $SO(D-p-3)$ isometry, the component of the metric corrections $H^{(T)}_{MN}$ along the isometry directions can be written as
\begin{equation}
H^{(T)}_{\theta_i\theta_j}=h l r_0^2 S_0^2 \Omega_{ij}.
\end{equation}
All other components of $H^{(T)}_{MN}$ with one component along $\theta_i$ directions is zero by the requirement of isometry. Keeping this in mind and using the property that $$P^{MN} H^{(T)}_{MN}=0,$$ we get
\begin{eqnarray}
h=-\frac{1}{D-p-3}\left(H^{(T)}_{MN}p^{MN}+\frac{H^{(T)}_{MN}z^M z^N}{(l r_0^2)^{-1}-n_S^2}\right).
\end{eqnarray}
The metric correction along the isometry direction in the effective spacetime language is $\frac{1}{D^2}\delta \phi^{(2)}$ and it is related in an obvious manner to the metric corrections in the geometric picture as 
\begin{eqnarray}
&&\frac{1}{D}\delta\phi^{(2)}=\frac{1}{D}H^{Tr}+ h\nonumber\\
\implies && H^{Tr}=\delta \phi^{(2)}+\frac{D}{D-p-3}\left(H^{(T)}_{MN}p^{MN}+\frac{H^{(T)}_{MN}z^M z^N}{(l r_0^2)^{-1}-n_S^2}\right)\nonumber\\
\implies && H^{Tr}=\mathcal{O}(1).
\end{eqnarray}
The metric correction in \eqref{geometric_met} in the tensor sector can be written as
\begin{eqnarray}
H^{(T)}_{MN}dx^M dx^N&=&H^{(T)}_{AB}P^A_M P^B_N dx^M dx^N\nonumber\\
&=& H^{(T)}_{AB}\left(p^A_M+\frac{z^A z_M}{(l r_0^2)^{-1}-n_S^2}+\Omega^A_M\right) \left(p^B_N+\frac{z^B z_N}{(l r_0^2)^{-1}-n_S^2}+\Omega^B_N\right) dx^M dx^N\nonumber\\
&=& H^{(T)}_{AB}\left(p^A_M p^B_N-\frac{p_{MN}}{p}p^{AB}\right) dx^M dx^N+2 z^BH^{(T)}_{AB}p^A_Mdx^M \frac{dz}{(l r_0^2)^{-1}-n_S^2}\nonumber\\
&&+H^{(T)}_{AB}z^A z^B\frac{dz^2}{((l r_0^2)^{-1}-n_S^2)^2}+h l r_0^2S_0^2 d\Omega^{D-p-3}+\frac{1}{p}H^{(T)}_{AB}p^{AB} p_{MN}dx^M dx^N.\nonumber\\
\end{eqnarray}
Also, \begin{eqnarray}
p_{MN}dx^M dx^N=lr_0^2dy^i dy^i.
\end{eqnarray}
This gives  the map of the tensor part in geometric picture to metric components in \eqref{effective_space_metric} as
\begin{eqnarray}
&&H^{(T)}_{AB}\left(p^A_M p^B_N-\frac{p_{MN}}{p}p^{AB}\right)=T_{MN}+\tilde{ \kappa} T_{MN,\tilde{ \kappa}}\quad, \frac{z^BH^{(T)}_{AB}p^A_M}{(l r_0^2)^{-1}-n_S^2}=V_{zi}+\tilde{ \kappa} V_{zi,\tilde{ \kappa}},\nonumber\\&&\quad  \frac{H^{(T)}_{AB}z^A z^B}{((l r_0^2)^{-1}-n_S^2)^2}=S_{zz}+\tilde{ \kappa}S_{zz,\tilde{ \kappa}}\quad \text{and}\quad H^{(T)}_{AB}p^{AB}=p (S_{tr}+\tilde{ \kappa}S_{tr,\tilde{ \kappa}}).\nonumber\\
\end{eqnarray}
Using this we can write 
\begin{eqnarray}
H^{Tr}=\delta \phi^{(2)}+p (S_{tr}+\tilde{ \kappa}S_{tr,\tilde{ \kappa}})+\left(S_{zz}+\tilde{ \kappa}S_{zz,\tilde{ \kappa}}\right)((l r_0^2)^{-1}-n_S^2).
\end{eqnarray}
In the last line we have used the maps mentioned above. 
\bibliographystyle{JHEP}
\bibliography{ssbib}

\providecommand{\href}[2]{#2}\begingroup\raggedright\begin{thebibliography}{10}

\bibitem{Emparan:2013moa}
R.~Emparan, R.~Suzuki and K.~Tanabe, \emph{{The large D limit of General
  Relativity}}, \href{https://doi.org/10.1007/JHEP06(2013)009}{\emph{JHEP}
  {\bfseries 1306} (2013) 009}
  [\href{https://arxiv.org/abs/1302.6382}{{\ttfamily 1302.6382}}].

\bibitem{Emparan:2013oza}
R.~Emparan and K.~Tanabe, \emph{{Holographic superconductivity in the large D
  expansion}}, \href{https://doi.org/10.1007/JHEP01(2014)145}{\emph{JHEP}
  {\bfseries 1401} (2014) 145}
  [\href{https://arxiv.org/abs/1312.1108}{{\ttfamily 1312.1108}}].

\bibitem{Emparan:2013xia}
R.~Emparan, D.~Grumiller and K.~Tanabe, \emph{{Large-D gravity and low-D
  strings}},
  \href{https://doi.org/10.1103/PhysRevLett.110.251102}{\emph{Phys.Rev.Lett.}
  {\bfseries 110} (2013) 251102}
  [\href{https://arxiv.org/abs/1303.1995}{{\ttfamily 1303.1995}}].

\bibitem{Emparan:2014cia}
R.~Emparan and K.~Tanabe, \emph{{Universal quasinormal modes of large D black
  holes}}, \href{https://doi.org/10.1103/PhysRevD.89.064028}{\emph{Phys.Rev.}
  {\bfseries D89} (2014) 064028}
  [\href{https://arxiv.org/abs/1401.1957}{{\ttfamily 1401.1957}}].

\bibitem{Emparan:2014jca}
R.~Emparan, R.~Suzuki and K.~Tanabe, \emph{{Instability of rotating black
  holes: large D analysis}},
  \href{https://doi.org/10.1007/JHEP06(2014)106}{\emph{JHEP} {\bfseries 1406}
  (2014) 106} [\href{https://arxiv.org/abs/1402.6215}{{\ttfamily 1402.6215}}].

\bibitem{Emparan:2014aba}
R.~Emparan, R.~Suzuki and K.~Tanabe, \emph{{Decoupling and non-decoupling
  dynamics of large $D$ black holes}},
  \href{https://doi.org/10.1007/JHEP07(2014)113}{\emph{JHEP} {\bfseries 1407}
  (2014) 113} [\href{https://arxiv.org/abs/1406.1258}{{\ttfamily 1406.1258}}].

\bibitem{Emparan:2015rva}
R.~Emparan, R.~Suzuki and K.~Tanabe, \emph{{Quasinormal modes of (Anti-)de
  Sitter black holes in the 1/D expansion}},
  \href{https://doi.org/10.1007/JHEP04(2015)085}{\emph{JHEP} {\bfseries 04}
  (2015) 085} [\href{https://arxiv.org/abs/1502.02820}{{\ttfamily
  1502.02820}}].

\bibitem{Emparan:2015hwa}
R.~Emparan, T.~Shiromizu, R.~Suzuki, K.~Tanabe and T.~Tanaka, \emph{{Effective
  theory of Black Holes in the 1/D expansion}},
  \href{https://doi.org/10.1007/JHEP06(2015)159}{\emph{JHEP} {\bfseries 06}
  (2015) 159} [\href{https://arxiv.org/abs/1504.06489}{{\ttfamily
  1504.06489}}].

\bibitem{Suzuki:2015iha}
R.~Suzuki and K.~Tanabe, \emph{{Stationary black holes: Large $D$ analysis}},
  \href{https://doi.org/10.1007/JHEP09(2015)193}{\emph{JHEP} {\bfseries 09}
  (2015) 193} [\href{https://arxiv.org/abs/1505.01282}{{\ttfamily 1505.01282}}].

\bibitem{Bhattacharyya:2015dva}
S.~Bhattacharyya, A.~De, S.~Minwalla, R.~Mohan and A.~Saha, \emph{{A membrane
  paradigm at large D}},
  \href{https://doi.org/10.1007/JHEP04(2016)076}{\emph{JHEP} {\bfseries 04}
  (2016) 076} [\href{https://arxiv.org/abs/1504.06613}{{\ttfamily
  1504.06613}}].

\bibitem{Bhattacharyya:2015fdk}
S.~Bhattacharyya, M.~Mandlik, S.~Minwalla and S.~Thakur, \emph{{A Charged
  Membrane Paradigm at Large D}},
  \href{https://doi.org/10.1007/JHEP04(2016)128}{\emph{JHEP} {\bfseries 04}
  (2016) 128} [\href{https://arxiv.org/abs/1511.03432}{{\ttfamily
  1511.03432}}].

\bibitem{Tanabe:2015isb}
K.~Tanabe, \emph{{Instability of the de Sitter Reissner–Nordstrom black hole
  in the $1/D$ expansion}},
  \href{https://doi.org/10.1088/0264-9381/33/12/125016}{\emph{Class. Quant.
  Grav.} {\bfseries 33} (2016) 125016}
  [\href{https://arxiv.org/abs/1511.06059}{{\ttfamily 1511.06059}}].

\bibitem{Bhattacharyya:2017hpj}
S.~Bhattacharyya, P.~Biswas, B.~Chakrabarty, Y.~Dandekar and A.~Dinda,
  \emph{{The large D black hole dynamics in AdS/dS backgrounds}},
  \href{https://doi.org/10.1007/JHEP10(2018)033}{\emph{JHEP} {\bfseries 10}
  (2018) 033} [\href{https://arxiv.org/abs/1704.06076}{{\ttfamily 1704.06076}}].

\bibitem{Bhattacharyya:2018szu}
S.~Bhattacharyya, P.~Biswas and Y.~Dandekar, \emph{{Black holes in presence of
  cosmological constant: Second order in $\frac{1}{D}$}},
  \href{https://doi.org/10.1007/JHEP10(2018)171}{\emph{JHEP} {\bfseries 10}
  (2018) 171} [\href{https://arxiv.org/abs/1805.00284}{{\ttfamily 1805.00284}}].

\bibitem{Kundu:2018dvx}
S.~Kundu and P.~Nandi, \emph{{Large D gravity and charged membrane dynamics
  with nonzero cosmological constant}},
  \href{https://doi.org/10.1007/JHEP12(2018)034}{\emph{JHEP} {\bfseries 12}
  (2018) 034} [\href{https://arxiv.org/abs/1806.08515}{{\ttfamily
  1806.08515}}].

\bibitem{Emparan:2015gva}
R.~Emparan, R.~Suzuki and K.~Tanabe, \emph{{Evolution and endpoint of the black
  string instability: Large D solution}},
  \href{https://doi.org/10.1103/PhysRevLett.115.091102}{\emph{Phys. Rev. Lett.}
  {\bfseries 115} (2015) 091102}
  [\href{https://arxiv.org/abs/1506.06772}{{\ttfamily 1506.06772}}].

\bibitem{Rozali:2016yhw}
M.~Rozali and A.~Vincart-Emard, \emph{{On Brane Instabilities in the Large $D$
  Limit}},
  \href{https://doi.org/10.1007/JHEP08(2016)166}{\emph{JHEP} {\bfseries 08}
  (2016) 166} [\href{https://arxiv.org/abs/1607.01747}{{\ttfamily 1607.01747}}].

\bibitem{Emparan:2016sjk}
R.~Emparan, K.~Izumi, R.~Luna, R.~Suzuki and K.~Tanabe, \emph{{Hydro-elastic
  Complementarity in Black Branes at large D}},
  \href{https://doi.org/10.1007/JHEP06(2016)117}{\emph{JHEP} {\bfseries 06}
  (2016) 117} [\href{https://arxiv.org/abs/1602.05752}{{\ttfamily
  1602.05752}}].

\bibitem{Bhattacharyya:2018iwt}
S.~Bhattacharyya, P.~Biswas and M.~Patra, \emph{{A leading-order comparison
  between fluid-gravity and membrane-gravity dualities}},
  \href{https://arxiv.org/abs/1807.05058}{{\ttfamily 1807.05058}}.

\bibitem{Bhattacharyya:2019mbz}
S.~Bhattacharyya, P.~Biswas, A.~Dinda and M.~Patra, \emph{{Fluid-gravity and
  membrane-gravity dualities - Comparison at subleading orders}},
  \href{https://arxiv.org/abs/1902.00854}{{\ttfamily 1902.00854}}.

\bibitem{Andrade:2018nsz}
T.~Andrade, R.~Emparan and D.~Licht, \emph{{Rotating black holes and black bars
  at large D}}, \href{https://doi.org/10.1007/JHEP09(2018)107}{\emph{JHEP}
  {\bfseries 09} (2018) 107}
  [\href{https://arxiv.org/abs/1807.01131}{{\ttfamily 1807.01131}}].

\bibitem{Andrade:2018rcx}
T.~Andrade, R.~Emparan and D.~Licht, \emph{{Charged rotating black holes in
  higher dimensions}},
  \href{https://doi.org/10.1007/JHEP02(2019)076}{\emph{JHEP} {\bfseries 02}
  (2019) 076} [\href{https://arxiv.org/abs/1810.06993}{{\ttfamily
  1810.06993}}].

\bibitem{Casalderrey-Solana:2018uag}
J.~Casalderrey-Solana, C.~P. Herzog and B.~Meiring, \emph{{Holographic Bjorken
  Flow at Large-$D$}},
  \href{https://doi.org/10.1007/JHEP01(2019)181}{\emph{JHEP} {\bfseries 01}
  (2019) 181} [\href{https://arxiv.org/abs/1810.02314}{{\ttfamily
  1810.02314}}].

\bibitem{Dandekar:2017aiv}
Y.~Dandekar, S.~Kundu, S.~Mazumdar, S.~Minwalla, A.~Mishra and A.~Saha,
  \emph{{An Action for and Hydrodynamics from the improved Large D membrane}},
  \href{https://doi.org/10.1007/JHEP09(2018)137}{\emph{JHEP} {\bfseries 09}
  (2018) 137} [\href{https://arxiv.org/abs/1712.09400}{{\ttfamily 1712.09400}}].

\bibitem{Dandekar:2016jrp}
Y.~Dandekar, S.~Mazumdar, S.~Minwalla and A.~Saha, \emph{{Unstable `black
  branes' from scaled membranes at large $D$}},
  \href{https://doi.org/10.1007/JHEP12(2016)140}{\emph{JHEP} {\bfseries 12}
  (2016) 140} [\href{https://arxiv.org/abs/1609.02912}{{\ttfamily
  1609.02912}}].

\bibitem{Bhattacharyya:2016nhn}
S.~Bhattacharyya, A.~K. Mandal, M.~Mandlik, U.~Mehta, S.~Minwalla, U.~Sharma
  et~al., \emph{{Currents and Radiation from the large $D$ Black Hole
  Membrane}}, \href{https://doi.org/10.1007/JHEP05(2017)098}{\emph{JHEP}
  {\bfseries 05} (2017) 098}
  [\href{https://arxiv.org/abs/1611.09310}{{\ttfamily 1611.09310}}].

\bibitem{Dandekar:2016fvw}
Y.~Dandekar, A.~De, S.~Mazumdar, S.~Minwalla and A.~Saha, \emph{{The large D
  black hole Membrane Paradigm at first subleading order}},
  \href{https://doi.org/10.1007/JHEP12(2016)113}{\emph{JHEP} {\bfseries 12}
  (2016) 113} [\href{https://arxiv.org/abs/1607.06475}{{\ttfamily 1607.06475}}].

\bibitem{Bhattacharjee:2015yaa}
S.~Bhattacharjee, S.~Sarkar and A.~C. Wall, \emph{{Holographic entropy
  increases in quadratic curvature gravity}},
  \href{https://doi.org/10.1103/PhysRevD.92.064006}{\emph{Phys. Rev.}
  {\bfseries D92} (2015) 064006}
  [\href{https://arxiv.org/abs/1504.04706}{{\ttfamily 1504.04706}}].

\bibitem{Bhattacharyya:2016xfs}
S.~Bhattacharyya, F.~M. Haehl, N.~Kundu, R.~Loganayagam and M.~Rangamani,
  \emph{{Towards a second law for Lovelock theories}},
  \href{https://doi.org/10.1007/JHEP03(2017)065}{\emph{JHEP} {\bfseries 03}
  (2017) 065} [\href{https://arxiv.org/abs/1612.04024}{{\ttfamily
  1612.04024}}].

\bibitem{Chen:2015fuf}
B.~Chen, Z.-Y. Fan, P.~Li and W.~Ye, \emph{{Quasinormal modes of Gauss-Bonnet
  black holes at large D}},
  \href{https://doi.org/10.1007/JHEP01(2016)085}{\emph{JHEP} {\bfseries 01}
  (2016) 085} [\href{https://arxiv.org/abs/1511.08706}{{\ttfamily
  1511.08706}}].

\bibitem{Chen:2017hwm}
B.~Chen and P.-C. Li, \emph{{Static Gauss-Bonnet Black Holes at Large $D$}},
  \href{https://doi.org/10.1007/JHEP05(2017)025}{\emph{JHEP} {\bfseries 05}
  (2017) 025} [\href{https://arxiv.org/abs/1703.06381}{{\ttfamily
  1703.06381}}].

\bibitem{Chen:2017rxa}
B.~Chen, P.-C. Li and C.-Y. Zhang, \emph{{Einstein-Gauss-Bonnet Black Strings
  at Large $D$}}, \href{https://doi.org/10.1007/JHEP10(2017)123}{\emph{JHEP}
  {\bfseries 10} (2017) 123}
  [\href{https://arxiv.org/abs/1707.09766}{{\ttfamily 1707.09766}}].

\bibitem{Chen:2018nbh}
B.~Chen, P.-C. Li, Y.~Tian and C.-Y. Zhang, \emph{{Holographic Turbulence in
  Einstein-Gauss-Bonnet Gravity at Large $D$}},
  \href{https://doi.org/10.1007/JHEP01(2019)156}{\emph{JHEP} {\bfseries 01}
  (2019) 156} [\href{https://arxiv.org/abs/1804.05182}{{\ttfamily
  1804.05182}}].

\bibitem{Saha:2018elg}
A.~Saha, \emph{{The large D Membrane Paradigm For Einstein-Gauss-Bonnet
  Gravity}}, \href{https://doi.org/10.1007/JHEP01(2019)028}{\emph{JHEP}
  {\bfseries 01} (2019) 028}
  [\href{https://arxiv.org/abs/1806.05201}{{\ttfamily 1806.05201}}].

\bibitem{Kats:2007mq}
Y.~Kats and P.~Petrov, \emph{{Effect of curvature squared corrections in AdS on
  the viscosity of the dual gauge theory}},
  \href{https://doi.org/10.1088/1126-6708/2009/01/044}{\emph{JHEP} {\bfseries
  01} (2009) 044} [\href{https://arxiv.org/abs/0712.0743}{{\ttfamily
  0712.0743}}].

\bibitem{Banerjee:2009wg}
N.~Banerjee and S.~Dutta, \emph{{Higher Derivative Corrections to Shear
  Viscosity from Graviton's Effective Coupling}},
  \href{https://doi.org/10.1088/1126-6708/2009/03/116}{\emph{JHEP} {\bfseries
  03} (2009) 116} [\href{https://arxiv.org/abs/0901.3848}{{\ttfamily
  0901.3848}}].

\bibitem{Andrade:2018zeb}
T.~Andrade, C.~Pantelidou and B.~Withers, \emph{{Large D holography with metric
  deformations}}, \href{https://doi.org/10.1007/JHEP09(2018)138}{\emph{JHEP}
  {\bfseries 09} (2018) 138}
  [\href{https://arxiv.org/abs/1806.00306}{{\ttfamily 1806.00306}}].

\end{thebibliography}\endgroup
\end{document}